\DocumentMetadata{}
\documentclass[acmsmall,screen]{acmart}

\usepackage{url}
\usepackage{pifont}
\usepackage{xspace}
\usepackage{amsthm}
\usepackage{booktabs}
\usepackage{hyperref}
\usepackage{makecell}
\usepackage{multirow}
\usepackage{graphicx}
\usepackage{colortbl}
\usepackage{subfigure}
\usepackage{enumerate}
\usepackage{todonotes}
\usepackage{tablefootnote}
\usepackage{algpseudocode}
\usepackage{color, xcolor}
\usepackage{amsmath, amsfonts}

\newcommand{\ie}{\textit{i.e.,}\xspace}
\newcommand{\eg}{\textit{e.g.,}\xspace}
\newcommand{\etc}{\textit{etc.}\xspace}
\newcommand{\etal}{\textit{et al.}\xspace}

\newcommand{\tabref}[1]{Table~\ref{#1}\xspace}
\newcommand{\figref}[1]{Fig.~\ref{#1}\xspace}
\newcommand{\secref}[1]{Section~\ref{#1}\xspace}
\newcommand{\equref}[1]{Equation~\ref{#1}\xspace}

\newcommand{\toolname}{{\sc PIRLTest}\xspace}
\newcommand{\mobapp}{20\xspace}
\newcommand{\webapp}{5\xspace}
\newcommand{\widgetype}{14\xspace}
\newcommand{\monkey}{Monkey\xspace}
\newcommand{\qtesting}{Q-testing\xspace}
\newcommand{\webexplor}{WebExplor\xspace}

\newcommand{\tabincell}[2]{\begin{tabular}{@{}#1@{}}#2\end{tabular}}

\setcopyright{acmcopyright}
\copyrightyear{2024}
\acmYear{2024}

\acmJournal{TOSEM}
\acmVolume{0}
\acmNumber{0}
\acmArticle{1}
\acmMonth{0}

\begin{document}

\title{Effective, Platform-Independent GUI Testing via Image Embedding and Reinforcement Learning}

\author{Shengcheng Yu}
\email{yusc@smail.nju.edu.cn}
\orcid{0000-0003-4640-8637}
\affiliation{
  \institution{State Key Laboratory for Novel Software Technology, Nanjing University}
  \city{Nanjing}
  \country{China}
  \postcode{210093}
}
\author{Chunrong Fang}
\email{fangchunrong@nju.edu.cn}
\authornote{\textbf{Chunrong Fang is the corresponding author.}}
\orcid{0000-0002-9930-7111}
\affiliation{
  \institution{State Key Laboratory for Novel Software Technology, Nanjing University}
  \city{Nanjing}
  \country{China}
  \postcode{210093}
}
\author{Xin Li}
\email{522022320075@smail.nju.edu.cn}
\orcid{0009-0001-7953-3920}
\affiliation{
  \institution{State Key Laboratory for Novel Software Technology, Nanjing University}
  \city{Nanjing}
  \country{China}
  \postcode{210093}
}
\author{Yuchen Ling}
\email{yuchen\_ling@smail.nju.edu.cn}
\orcid{0009-0006-9227-3824}
\affiliation{
  \institution{State Key Laboratory for Novel Software Technology, Nanjing University}
  \city{Nanjing}
  \country{China}
  \postcode{210093}
}
\author{Zhenyu Chen}
\email{zychen@nju.edu.cn}
\orcid{0000-0002-9592-7022}
\affiliation{
  \institution{State Key Laboratory for Novel Software Technology, Nanjing University}
  \city{Nanjing}
  \country{China}
  \postcode{210093}
}
\author{Zhendong Su}
\email{zhendong.su@inf.ethz.ch}
\orcid{0000-0002-2970-1391}
\affiliation{
  \institution{Department of Computer Science, ETH Zurich}
  \city{Zurich}
  \country{Switzerland}
  \postcode{8092}
}

\begin{abstract}

Software applications (apps) have been playing an increasingly important role in various aspects of society. In particular, mobile apps and web apps are the most prevalent among all applications and are widely used in various industries as well as in people's daily lives. To help ensure mobile and web app quality, many approaches have been introduced to improve app GUI testing via automated exploration, including random testing, model-based testing, learning-based testing, \etc Despite the extensive effort, existing approaches are still limited in reaching high code coverage, constructing high-quality models, and being generally applicable. Reinforcement learning-based approaches, as a group of representative and advanced approaches for automated GUI exploration testing, are faced with difficult challenges, including effective app state abstraction, reward function design, \etc Moreover, they heavily depend on the specific execution platforms (\ie Android or Web), thus leading to poor generalizability and being unable to adapt to different platforms. 

This work specifically tackles these challenges based on the high-level observation that apps from distinct platforms share commonalities in GUI design. Indeed, we propose \toolname, an effective platform-independent approach for app testing. Specifically, \toolname utilizes computer vision and reinforcement learning techniques in a novel, synergistic manner for automated testing. It extracts the GUI widgets from GUI pages and characterizes the corresponding GUI layouts, embedding the GUI pages as states. The app GUI state combines the macroscopic perspective (app GUI layout) and the microscopic perspective (app GUI widget), and attaches the critical semantic information from GUI images. This enables \toolname to be platform-independent and makes the testing approach generally applicable on different platforms. \toolname explores apps with the guidance of a curiosity-driven strategy, which uses a Q-network to estimate the values of specific state-action pairs to encourage more exploration in uncovered pages without platform dependency. The exploration will be assigned with rewards for all actions, which are designed considering both the app GUI states and the concrete widgets, to help the framework explore more uncovered pages. We conduct an empirical study on \mobapp mobile apps and \webapp web apps, and the results show that \toolname is zero-cost when being adapted to different platforms, and can perform better than the baselines, covering 6.3--41.4\% more code on mobile apps and 1.5--51.1\% more code on web apps. \toolname is capable of detecting 128 unique bugs on mobile and web apps, including ~100 bugs that cannot be detected by the baselines.

\end{abstract}



\maketitle

\section{Introduction}

Testing is an effective method to ensure software quality. Current approaches utilize different strategies to automatically test the applications (app)\footnote{``Software'' and ``application'' are used interchangeably in this paper.}\cite{choudhary2015automated}, but such approaches still leave quite room for improvement. Manually written test scripts based on specific frameworks \cite{appium2022appium, selenium2022selenium} are commonly used in app testing \cite{gomez2013reran, guo2019sara, halpern2015mosaic, yu2021layout}. Testers are required to identify the target widgets and assign the test events. However, the manual work is quite a waste of time and human resources, creating a great burden on app developers. The involvement of manual effort of testers with different expertise makes test scripts' quality ranges widely.

One most widely used automated testing strategy is the random-based exploration \cite{google2022monkey, machiry2013dynodroid}, which can generate pseudo-random test events to fuzz the apps with relatively high efficiency brought by the random feature. However, most of the generated test events are invalid, which may not trigger interactive responses. \monkey \cite{google2022monkey}, as a representative tool, may generate test events like clicking on blank areas, inputting texts into \texttt{ImageView} widgets, \etc To increase the exploration effectiveness, model-based strategy \cite{athaiya2017testing, biagiola2019diversity, mao2016sapienz, mesbah2012crawling, su2017guided} is adopted to construct models for apps under test with dynamic or static app analysis technologies. Two main factors, the constructed model itself and the exploration strategy, determine the effectiveness of the model-based approaches. Due to the limitations of existing app analysis technologies, the generated models cannot reach high code coverage, thus imposing restrictions on the model-based strategy. Besides, specific app states require domain knowledge to generate valid inputs, which is tough for existing exploration strategies of automated testing. Learning-based technologies have increasing adaptive applications in automated testing \cite{adamo2018reinforcement, collins2021deep, pan2020reinforcement, romdhana2021deep, zheng2021automatic}. Reinforcement learning (RL) \cite{kaelbling1996reinforcement, watkins1992q} is an effective algorithm to help effectively explore the app states. \tabref{tab:rw} shows the comparison of representative RL-based approaches and \toolname. The ``Platform'' column show the platform supporting situation. The ``State Abstraction'' column shows the basis for state abstraction. The ``Supported Action Type'' column shows the support action numbers on mobile and web platforms, respectively. The ``Exploration Strategy'' column shows how the explored states and corresponding reward values are stored and utilized for further exploration guidance. The ``Reward Function Basis'' column shows the important factors considered in the reward function design.

\begin{table}[!htbp]
\centering
\caption{Comparison among Representative RL-based Approaches. 
\textit{Platform}: support for mobile or web platforms. 
\textit{State Abstraction}: support for mobile or web platforms. 
\textit{Supported Action Type}: supported action numbers of different types on mobile and web platforms. 
\textit{Exploration Strategy}: Storage and utilization of explored states and corresponding reward values for further exploration guidance. 
\textit{Reward Function Basis}: factors considered in the reward function design (closely related to \textit{State Abstraction}). 
}

\scalebox{0.7}{
\begin{tabular}{ccccccccc}

\toprule
\multirow{2}{*}{Approach} & 
\multicolumn{2}{c}{Platform} & 
\multirow{2}{*}{\tabincell{c}{State \\ Abstraction}} & 
\multicolumn{3}{c}{\tabincell{c}{Supported Action Type \\ (mobile|web)}} & 
\multirow{2}{*}{Exploration Strategy} & 
\multirow{2}{*}{Reward Function Basis}
\\ \cmidrule{2-3} \cmidrule{5-7}
                          
& Mobile & Web & & Widget & Page & System & & \\ \midrule

QBE \cite{koroglu2018qbe}             & \checkmark & & Layout File      & 5|0 & 1|0 & 5|0 & Q-table        & \textit{Not Specified} \\ 
Q-testing \cite{pan2020reinforcement} & \checkmark & & Layout File      & 4|0 & 2|0 & 3|0 & Q-table        & Activity Comparison \\ 
WebExplor \cite{zheng2021automatic}   & & \checkmark & HTML Doc         & 0|1 & 0|0 & 0|0 & Q-function     & State-Action-State Tuple \\ 
ARES \cite{romdhana2021deep}          & \checkmark & & Layout File      & 4|0 & 0|0 & 2|0 & Neural Network & Action Comparison \\ 
Fastbot2 \cite{lv2022fastbot2}        & \checkmark & & Layout File      & 2|0 & 0|0 & 0|0 & Q-table        & Action Comparison \\ 
DinoDroid \cite{zhao2022dinodroid}    & \checkmark & & Event Flow       & 4|0 & 1|0 & 2|0 & DQN            & Code Coverage \& Bug Detection \\ \midrule
\textbf{PIRLTest}  & \checkmark & \checkmark   & \tabincell{c}{\textbf{GUI Widget} \\ \textbf{\& GUI Layout}} & \textbf{5|6} & \textbf{3|3} & \textbf{6|5} & \textbf{DQN}        & \tabincell{c}{\textbf{GUI State Transition Times \&} \\ \textbf{GUI Widget Exploration Rate}} \\ \bottomrule
\end{tabular}
}
\label{tab:rw}
\end{table} 

However, it is still a tricky problem to effectively ensure the app quality due to its large exploration space. Even if it is shown in many studies that the RL framework is an advanced solution to app testing, current RL-based approaches are still faced with many problems. Existing approaches rely heavily on the platform features or interfaces, which prevents them from being adapted across different platforms with an effective abstraction to the app state. For example, \qtesting \cite{pan2020reinforcement} utilizes the layout files of Android apps as the app states; \webexplor \cite{zheng2021automatic} utilizes the front page HTML code as the app states. This brings much extra overhead when developers conduct testing on different platforms or they migrate tests among different platforms. Moreover, app state abstraction is an important basis for the reward function design, which assigns different reward values to actions that make covered or uncovered pages. However, according to our investigation and existing work \cite{chen2020object, yu2021layout}, we find that apps of different platforms adopt highly similar widget styles\footnote{Customized rendering is not considered.} (\eg \texttt{Button}, \texttt{TextField}, \texttt{ImageView}, \etc). This suggests that the GUI images can be utilized as app states to guide the RL-based strategy to make the exploration independently adaptive to apps of different platforms. Another problem of app state abstraction of current automated testing approaches is widget identification. Current approaches rely on the layout files to identify the target widgets \cite{pan2020reinforcement}. However, most apps adopt a hybrid version and embed \texttt{Canvas} elements or customized elements into the app pages. Such GUI widgets cannot be identified with existing approaches. As is often the case, a group of widgets will be considered as only one single widget, which greatly affects the exploration effectiveness. Consequently, identifying widgets directly from the app GUI is a good choice because it views the app GUI just like app users.

In order to dispose of the aforementioned challenges, we propose a novel approach, named \toolname, to realize \underline{\textbf{P}}latform-\underline{\textbf{I}}ndependent GUI \underline{\textbf{Test}}ing via image embedding and \underline{\textbf{R}}einforcement \underline{\textbf{L}}earning. We adopt a deep Q-network (DQN) \cite{mnih2013playing} to guide \toolname to explore more app states. First, we propose a novel algorithm to characterize the app state from the perspective of GUI to avoid dependency on platform features. The accuracy of the state characterization of an app can significantly affect the exploration capabilities of the RL framework. Specifically, we combine a microscopic perspective (\ie widget feature) and a macroscopic perspective (\ie structure feature) to complete the app GUI embedding. For the widget feature, we extract all the widget images and their coordinates with the computer vision (CV) algorithms. The embedded widget feature is composed of three parts. First, the widget image is embedded in a vector with a convolutional neural network (CNN); second, the widget coordinate is fed into a CNN to generate a vector after the corresponding pixels of the widget in the app page to black and other pixels to white; the third part of the widget feature is a \widgetype-length one-hot vector representing the widget type recognized by a CNN model. Then, the three vectors are concatenated to represent a widget, and we get the average vector of all widget features. For the structure feature, we use a tree structure to represent the layout relationships of all the widgets, neglecting the concrete widget images. The trees are transformed into strings with curly braces to keep the relative relationships. The strings are fed into a recurrent neural network (RNN) to obtain a vector. The widget feature and the structure feature are concatenated to accomplish the app state (\ie page) embedding.

\toolname adopts the curiosity-driven \cite{burda2018large} strategy in the DQN algorithm to explore more app pages with the test event (\ie action) execution. Specifically, given the app state, executable actions are inferred, then the Q-network of \toolname, trained with the history exploration data stored in the exploration memory, is used to determine the values of all state-action pairs. The values of state-action pairs are transferred to the execution probabilities based on the Boltzmann strategy \cite{cesa2017boltzmann} to avoid the ``exploration-exploitation dilemma'' \cite{robbins1952some}, and the state-action pair with the highest probability is executed. The reward function is a critical part of the RL framework to train the Q-network to better guide the algorithm to explore more uncovered pages. The design of the reward function in \toolname is based on the exploration rate of one specific app page and the page transition times. Such a design relies heavily on the app GUI state representation with a combination of the macroscopic perspective (app GUI layout) and the microscopic perspective (app GUI widget). Exploration rate refers to the explored action percentage in specific app pages, and page transition times can tell how many times this page is transited to. Both are based on the page similarity with the app GUI image embedding.

We conduct an empirical study to evaluate \toolname on different platforms. The experiment results show that \toolname achieves satisfying effectiveness on different platforms, with no extra modification or changes to generate tests on different platforms. \toolname covers 6.3--41.4\% more code on mobile apps and 1.5--51.1\% more code on web apps, and it covers a large percentage of code that is not covered by the baselines. \toolname is capable of detecting 128 unique bugs, including ~100 ones that cannot be detected by the baselines.

We declare the following noteworthy contributions of this paper: 

\begin{itemize}
  \item We introduce a platform-independent GUI testing approach via image embedding and reinforcement learning, which is zero-cost when being adapted to different platforms.
  \item We propose a novel algorithm to abstract the app GUI states with the GUI image embedding, including the widget extraction (microscopic perspective) and the layout characterization (macroscopic perspective). This algorithm helps effectively characterize the \textit{State}s in the RL model.
  \item We propose a novel reward function design in the RL framework, with a comprehensive consideration of the exploration of both app GUI as a whole and concrete GUI widgets.
  \item We implement a tool and conduct an empirical evaluation of the effectiveness of \toolname on both mobile apps and web apps, which shows the outstanding performance of \toolname over the representative baselines.
\end{itemize}

\textbf{More information and the reproduction package is available on \url{https://sites.google.com/view/pirltest}.}

\section{Background \& Motivation}

In this section, we present the background and the motivation of this work, including the preliminaries of reinforcement learning, and the limitations of current approaches. We present the commonalities of mobile apps and web apps as the intuition to complete this paper.

\subsection{Deep Reinforcement Learning}

\begin{figure}[!htbp]
    \centering
    \includegraphics[width=0.5\linewidth]{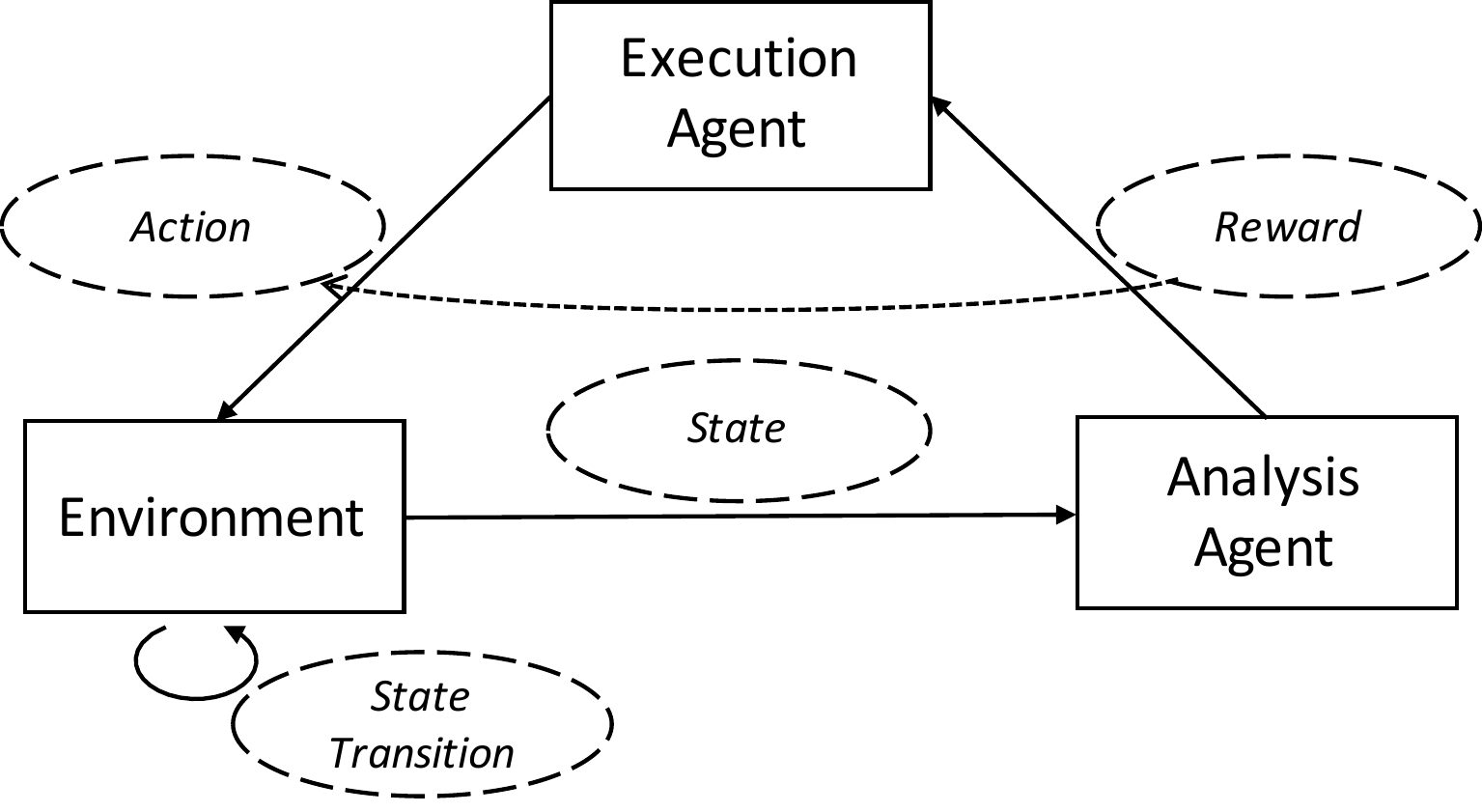}
    \caption{Markov Decision Process (MDP)}
    \label{fig:mdp}
\end{figure}

Reinforcement learning is a group of machine learning algorithms that make decisions and change behaviors based on environmental feedback to maximize expected benefits \cite{kaelbling1996reinforcement}. Reinforcement learning can be modeled as a Markov decision process \cite{white1989markov} (MDP) as shown in \figref{fig:mdp}. The MDP can be formalized as a four-tuple $ \langle \mathcal{S}, \mathcal{A},\mathcal{R}, \mathcal{P} \rangle $, where $\mathcal{S}$ refers to the set of all the states, $\mathcal{A}$ refers to the set of all the actions, $\mathcal{R}$ refers to the reward function: $\mathcal{S} \times \mathcal{A} \rightarrow \mathcal{R}$, and $\mathcal{P}$ refers to the transition probability function: $P\left(s, a, s^{\prime}\right) = P\left(s_{t+1}=s^{\prime} \mid s_{t}=s, a_{t}=a\right)$. In the scenario of automated testing, state means the current situation of the app under test, like the widgets and their contents presented on the app GUI, and the corresponding GUI layout, \ie relationships among different GUI widgets. Action means the events that can lead to the change of app GUI, like the activity transition. Generally speaking, at each timestamp $t$, the agent senses the state $s_t$ from the environment and then selects an action $a_t$ based on a specific policy $\pi \left( a|s \right)$. As the $a_t$ is executed, a state transition to a new state $s_{t+1}$ will happen, and a reward $r_t$ should be assigned. The object of RL is to train such an agent to maximize the expected discounted cumulative reward.

When traditional tabular RL algorithms are faced with the high-dimension or unbounded state and action spaces, it is impractical and unstable to learn an efficient policy merely based on such tabular algorithms. Therefore, a group of more robust RL algorithms combining the deep neural networks, named deep RL, is proposed to overcome the shortcomings. Deep Q-network \cite{mnih2013playing} is one model-free representative. In DQN, a neural network is used to approximate the state-action value function $Q^\pi$. The neural network is trained to estimate the expected discounted cumulative reward under the optimal policy $\pi^*$. With the combination of reinforcement learning and deep learning, \ie the DQN, it will become more effective to explore the environment.

\subsection{Commonalities of Mobile \& Web Apps}

In order to apply the \toolname on different platforms, including web platform and mobile platform, one significant prerequisite is the commonalities in app GUI of mobile apps and web apps. The intuition of \toolname is to extract all the widgets from the app pages for the exploration of the RL algorithm (details in \secref{sec:state}). The app pages are composed of widgets. Besides, the widgets used in mobile apps\footnote{\url{https://developer.android.com/reference/android/view/View}} and in web apps\footnote{\url{https://www.w3.org/}} can be mapped. We list several widely used widgets and their mapping relationships. As is depicted in \tabref{tab:commonality}, the Android widget classes correspond to specific HTML tags. From the visual aspect, the widget pairs share universal appearances.

\begin{table}[!htbp]
\centering
\caption{Commonalities of Mobile App \& Web App}
\scalebox{1}{
\begin{tabular}{c|c}
	\toprule
	Android Class & Web (HTML) Tag            \\ \midrule
	Button        & $<$button$>$                  \\
	TextView      & $<$p$>$                       \\
	EditText      & $<$input$>$                   \\
	CheckBox.     & $<$input type=``checkbox''$>$ \\
	ImageButton   & $<$input type=``image''$>$    \\
	ImageView     & $<$img$>$                     \\
	RadioButton   & $<$input type=``radio''$>$    \\
	Switch        & $<$input type=``checkbox''$>$ \\
	SeekBar       & $<$input type=``range''$>$    \\
	ProgressBar   & $<$progress$>$                \\ \bottomrule
\end{tabular}}
\label{tab:commonality}
\end{table}

Moreover, corresponding widgets share similar applicable actions (details in \secref{sec:action}). For example, for a clickable \texttt{Button} in a mobile or a web app, the possible actions are \textit{click}, \textit{right-click}, \etc; for an editable \texttt{TextField} (or \texttt{<input>} tag), the possible actions are \textit{input}, \textit{long-click}, \etc

Considering the above commonalities between widgets of mobile apps and web apps, and starting both from the visual aspect and the functionality correspondence, we can develop an effective approach that utilizes the GUI information with the help of CV technologies to guide the exploration of the reinforcement learning algorithm.

The commonalities of apps on different platforms inspire us to develop the platform-independent automated testing approach. First, from the user perspective, although apps on different platforms may have differences in their implementation, they tend to have similar GUI designs and GUI layouts \cite{yu2021layout}. It is nature for app users to perceive the apps with a platform-independent cognition. Apps on different platforms also adopt similar widget types and interactions from the user perspective. The RL algorithm is designed to simulated the testing process of app users to explore different activities in the apps. Therefore, it is reasonable to design the platform-independent automated testing approach from the actual user perspective. Second, from the app developer perspective, they always need to develop different versions on different platforms to attract more users with different devices. It forms a bar for app developers to use different tools for the automated testing for different platforms. As far as we know, some commercial companies hope to use a universal tool on all platforms instead of individual tools on different platforms, but none of existing tools can support this. In other fields of automated testing, researchers are also developing platform-independent tools, like the Appium\footnote{\url{https://appium.io/}}, one of the most popular tools to assist develop test scripts on different platforms, with which testers still need to maintain scripts with the same structure but different concrete contents (\ie widget identifiers). LIRAT by Yu \etal \cite{yu2021layout} is a fully platform-independent GUI test script record and replay tool that uses the visual-based app GUI widget matching. Therefore, we believe our work is significant as it is the first platform-independent exploration testing work that can be applied on different platforms.

\subsection{Limitations of Current Approaches}
\label{sec:challenge}

For current approaches that apply RL algorithms in app testing, like \cite{pan2020reinforcement, romdhana2021deep, zheng2021automatic, zheng2019wuji}, almost all of them depend on specific tools to capture the GUI structures. For example, UIAutomator\footnote{\url{https://developer.android.com/training/testing/other-components/ui-automator}} is a tool that can extract the mobile app page elements and the structure. \webexplor obtains the HTML elements from the browser. However, one prominent problem is that more and more pages adopt a hybrid version or have a highly customized widget due to the deeper customization of app GUIs. Under such circumstances, these widgets are hardly possible to be obtained. 

\begin{figure}[!htbp]
  \subfigure[Mobile App]{
  \begin{minipage}[t]{0.45\linewidth}
   \label{fig:moblimit}
   \centering
   \includegraphics[width=\linewidth]{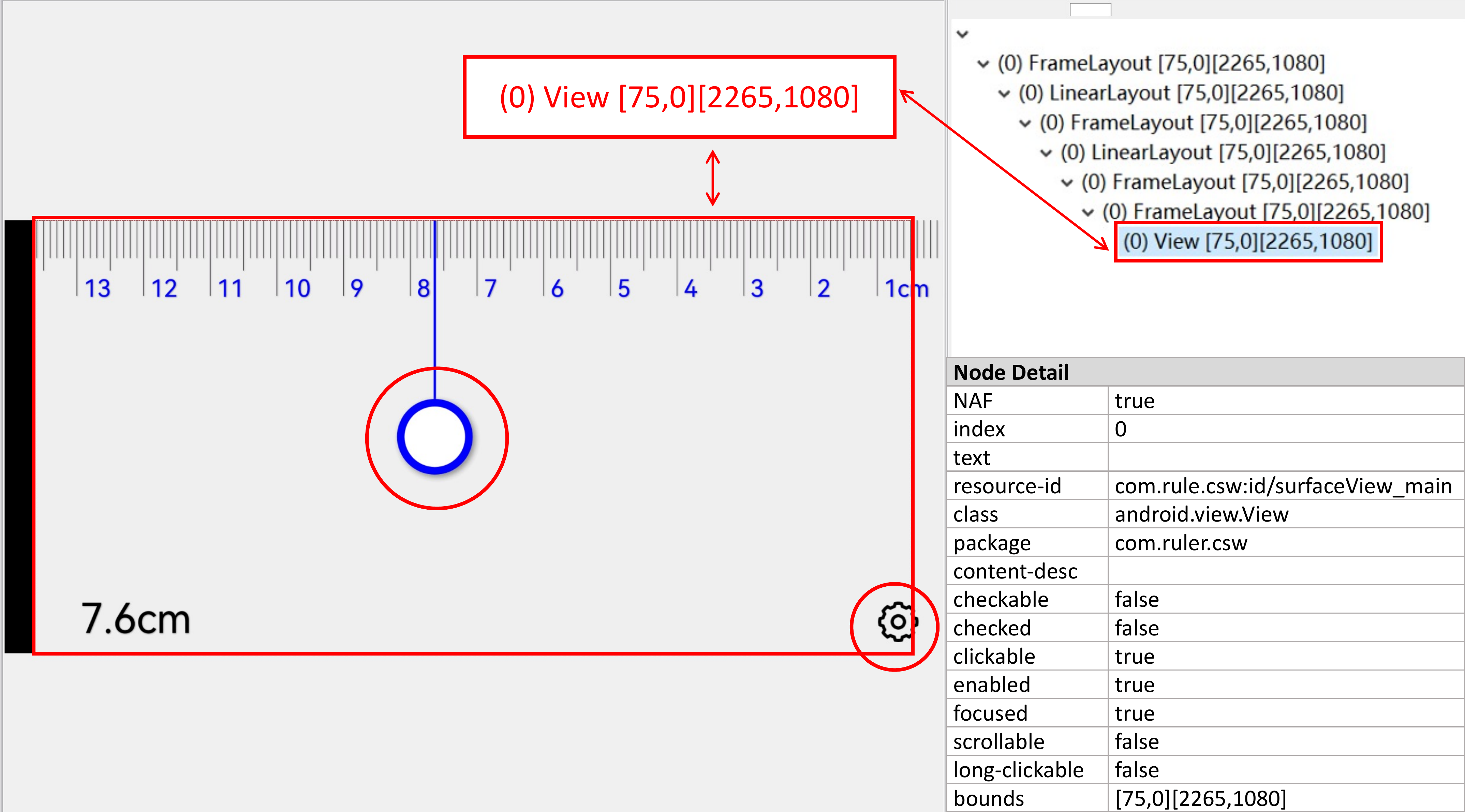}
  \end{minipage}
 }
 \hfill
 \subfigure[Web App]{
  \begin{minipage}[t]{0.45\linewidth}
   \label{fig:weblimit}
   \centering
   \includegraphics[width=\linewidth]{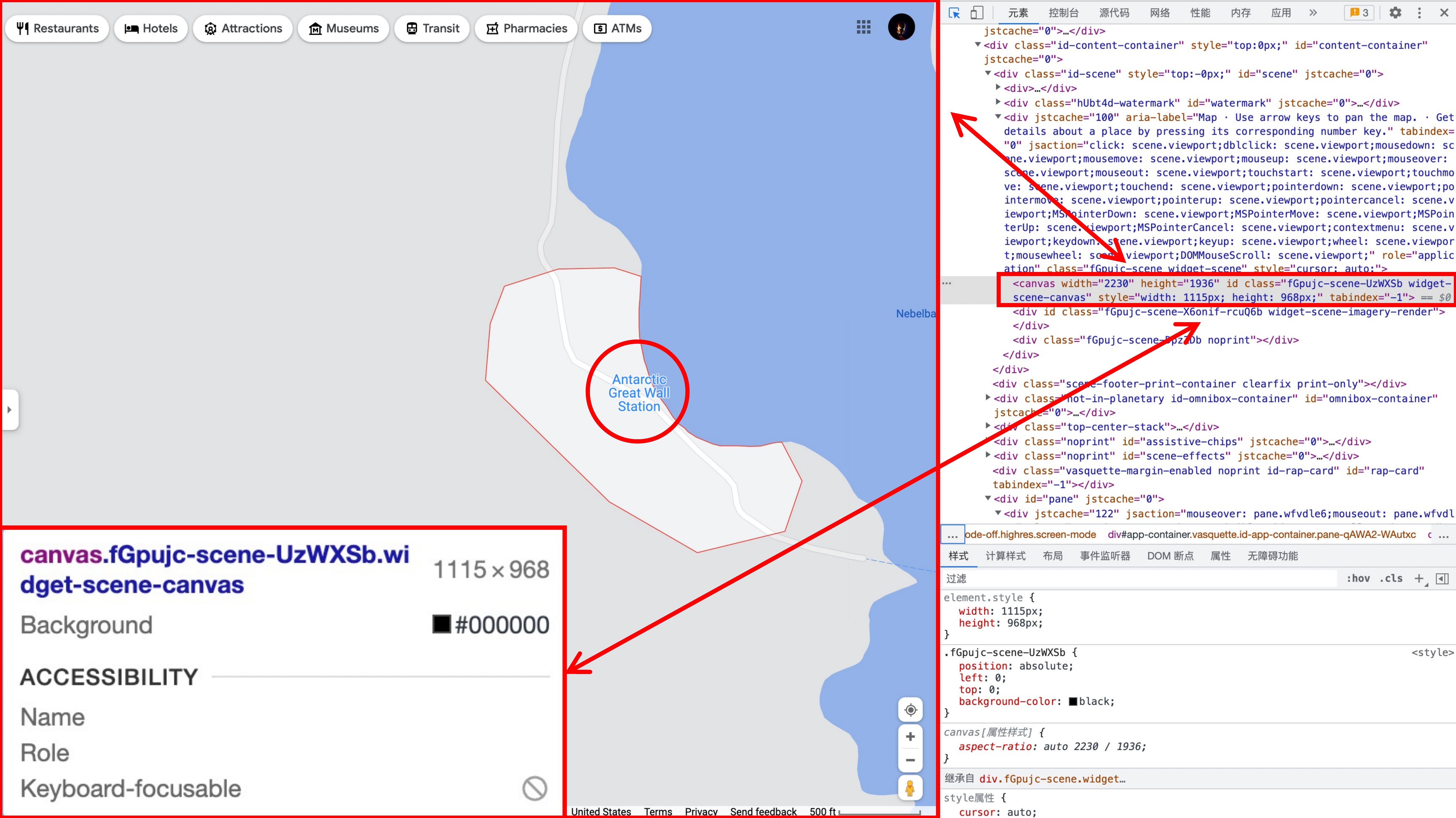}
  \end{minipage}
 }
 \centering
 \caption{Examples of the limitations of current approaches (mobile and web apps)}
\label{fig:exp}
\end{figure}

\figref{fig:moblimit} shows an example of a mobile app. In this \textit{Rule} app, the rule is a customized widget. The blue-edge widget can be dragged, and the gear-like button can be clicked. However, owing that this widget is a customized one, the rule is recognized as a whole widget in the class \texttt{android.view.View}. Also, the blue-edge widget and the gear-like button cannot be obtained by program analysis tool because they are just images shown on the \texttt{Canvas} widget, which may negatively affect the app exploration effectiveness.

\figref{fig:weblimit} gives an example from the web version of Google Maps. The red-circled texts are clickable and will trigger an information page, showing the details about the location. However, texts of location names are not accessible from the HTML layout structure files given that the whole page is a simple \texttt{Canvas} widget from the perspective of the HTML files.

For \toolname, which solves the limitations by utilizing the computer vision technologies, the widgets mentioned above can be identified without relying on the underlying layout files. More details are introduced in \secref{sec:widget} and \secref{sec:layout}.
 
\section{Approach} 

\begin{figure}[!htbp]
    \centering
    \includegraphics[width=\linewidth]{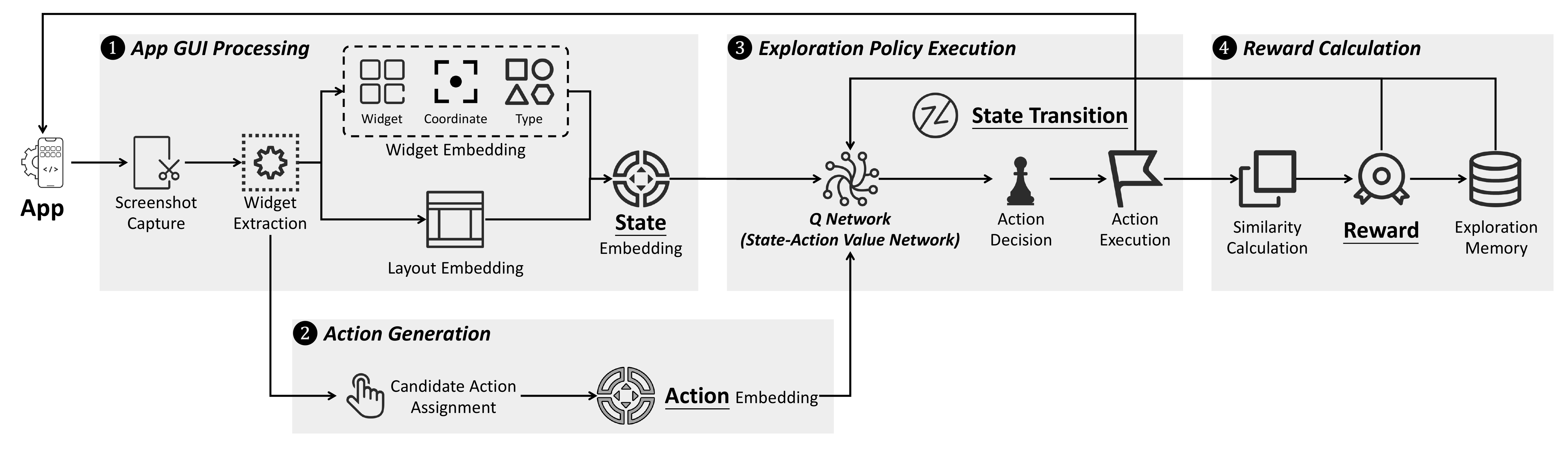}
    \caption{\toolname Framework}
    \label{fig:framework}
\end{figure}

In this section, we present the detailed design of \toolname. The app GUI screenshots are processed with CV algorithms to embed the app page into a vector so as to represent the states (\figref{fig:framework} \ding{202}), combining the macroscopic perspective (app GUI layout) and the microscopic perspective (app GUI widget). Based on the extracted widgets during the app GUI processing, candidate actions are generated and assigned to the corresponding widgets (\figref{fig:framework} \ding{203}), then \toolname embed the actions into vectors. Afterwards, \toolname calculates the state-action values with the Q-network and assigns different probabilities to the actions according to the values (\figref{fig:framework} \ding{204}). The Q-network is trained with the data from explored states, which are stored in the memory, as the curiosity-driven strategy will encourage \toolname to explore different states based on the reward calculation (\figref{fig:framework} \ding{205}). In the following sections, we will give a detailed depiction of the \toolname design.

\subsection{App GUI Processing} 
\label{sec:state}

One of the most significant parts of an RL framework is to abstract the state, which is a prerequisite for the reward calculation to reward the new state exploration. Traditional learning-based approaches tend to dump the XML layout files to abstract app state, which is quite inaccurate according to the challenges mentioned in \secref{sec:challenge}. Therefore, we start the app state embedding directly from the GUI images. App GUI (activity) is composed of different types of GUI widgets, which are the basic elements. The layout of such widgets, \ie the arrangements, describes the basic skeleton of the app GUI. Therefore, we focus on the GUI widgets and the widget layout during the state embedding. The widget layout describes the structure of the whole app GUI from the macroscopic perspective, and the widgets themselves describe the concrete contents of the app GUI from the microscopic perspective. Combining the two perspectives can help more precisely embed the app GUI images. The following two sections describe the detailed processing of widget extraction \& embedding and layout characterization \& embedding.

\subsubsection{\textbf{GUI Widget Extraction \& Embedding}}
\label{sec:widget}

The first step of GUI processing is to extract GUI widgets from the app screenshots, widgets are the basic components of app GUI images and are also the prerequisite for layout characterization. We adopt the classic computer vision algorithm, the Canny \cite{canny1986computational} edge detection, which is widely used in widget recognition in GUI analysis \cite{chen2020object}, to our approach. Targeting at one specific app screenshot of being turned grayscale, the Canny algorithm identifies the edges of widgets. Then the edges will be extracted to form an image of binary thresholding, where edges are represented with white pixels and non-edges with black pixels. Afterward, the edges of the widgets are obtained. However, there are still some noise data to be processed, \eg the misrecognized subtle areas. Such noise data will not be assigned with interactable attributes, so we have to eliminate such areas to avoid misleading during the RL process. The accordance is the size ratio of the widget to the whole screenshot, and we set a threshold, following existing studies \cite{yu2021layout, chen2020object}, to eliminate the subtle misrecognized areas.

We use the IoU metric to evaluate the effectiveness, which is a widely used metric \cite{chen2020object} for object detection and means the overlap between the detected widget areas and the labeled ground-truth widget areas. We set the widgets with IoU bigger than 0.8 as true positive (the threshold is set according to existing studies \cite{chen2020object, yu2021layout}). The larger IoU is, the more accurate the detected results are. In the evaluation, true positive refers to a detected widget which matches a ground truth widget. False positive refers to a detected widget which does not match any ground truth widgets. False negative refers to a ground truth widget which is not matched by any detected boxes. We first construct an evaluation dataset. We invite experts to label and cross-validate 17934 widgets on 1000 GUI screenshots randomly collected from different categories of apps (no overlap with our experiment subjects). Then, we run the Canny algorithm on the GUI screenshots to evaluate the effectiveness. The precision value is 94.73\%, the recall value is 99.09\%, the F-measure value is 96.86\%, and the accuracy is 93.92\%. The results reflect the high effectiveness of the Canny algorithm in our scenario in detecting GUI widgets from the visual perspective.

All obtained single widgets view the app GUI from a microscopic perspective. We consider three attributes of a widget image for each widget in the app GUI: the widget image, the widget location, and the widget type. These three attributes are combined to determine two widgets on different app GUI images are the same or similar widgets.

\textbf{\textit{\underline{Widget Image}}} refers to the widget extracted from the GUI screenshot. For each widget, we feed the widget image into a convolutional neural network to embed it into a 4096-dimension vector. Specifically, the CNN we use here is a VGG-16 model \cite{simonyan2014very}. The model is a pre-trained model in \cite{yu2021prioritize} with the purpose of classifying widget types. The model is trained with a dataset containing over 36k widget images manually collected, labeled, and verified by the authors. In order to obtain the embedding vector, we delete the last layer of the model, which outputs the probabilities indicating the types. The last but one layer is a \texttt{FullyConnected} layer containing 4096 neurons, and it will output a 4096-dimension vector, which can be viewed to represent the features of the widget images \cite{li2019boosting, jang2020approach}.

\textbf{\textit{\underline{Widget Location}}} indicates the location of the widget within the page. Traditionally, in order to represent the coordinate of a widget, common practices use four integers: the coordinate of the left-upper corner and the width and height of the widget. The coordinates can be obtained during the widget extraction. However, if we directly use the four integers to embed location information, the weight of location to the final widget feature is too low and will be neglected by the model. Therefore, we use a black-and-white GUI image to represent the widget location. We turn the widget part of a GUI image into all-white and the other part into all-black. Then, the transformed GUI image is fed into the VGG-16 model mentioned above to obtain a 4096-dimension vector to represent the location features, which is the same practice on widget images.

\textbf{\textit{\underline{Widget Type}}} is identified by a convolutional neural network. Widget type has a close relationship to the applicable actions, and the type attribute will affect the exploration. Thus, we consider embedding the widget type. As mentioned in the \textbf{\textit{\underline{Widget Image}}} part, the VGG-16 model is trained with the purpose of identifying the widget type. The model covers 14 widely used widgets \cite{yu2021prioritize} which are applicable to different platforms. We use a 14-dimension one-hot vector to represent the widget type. Compared with the widget image and the widget location, the importance of the widget type is weaker, so we do not expand the dimensions. We evaluate the classification results on 14 categories with precision, recall, F-measure, and accuracy. The results are 90.05\%, 89.97\%, 90.01\%, and 89.98\%, respectively. The results show the good performance of the classification results.

The three kinds of widget embedding results are concatenated to form the widget representation vector, with a length of 8206 dimensions ($2 \times 4096 + 14$). The 8206-dimension vector is the embedding representation of one single widget on the app GUI.

\subsubsection{\textbf{GUI Layout Characterization \& Embedding}}
\label{sec:layout}

\begin{figure}[!htbp]
    \centering
    \includegraphics[width=0.8\linewidth]{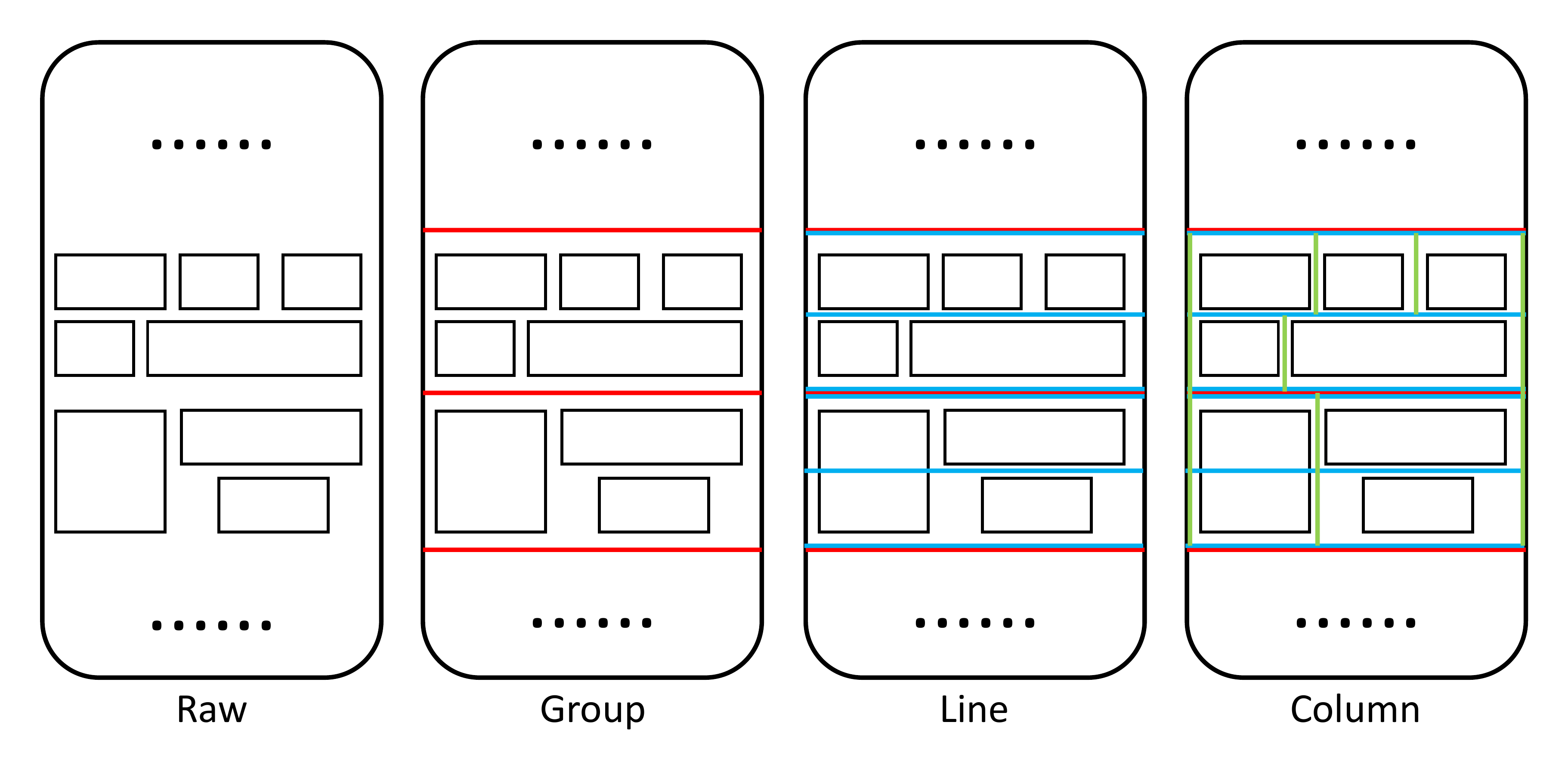}
    \caption{Layout Characterization of App GUI}
    \label{fig:characterization}
\end{figure}

After obtaining the widgets from the app screenshots, the layout of the app screenshots should be characterized, which illustrates the overall structure of the screenshot and the relative positions among all widgets. Beyond concrete widgets, layout characterization effectively reflects the whole panorama of the GUI structure, which is important in state determination. During the GUI design, the designers tend to arrange the widgets with close semantics together, and arrange widgets with different ones apart \cite{seebach2001cranky, raymond2003applying, li2022cross}. Consequently, we characterize the screenshot layout with a four-level hierarchy (\figref{fig:characterization}), the whole app GUI, the \texttt{Group}, the \texttt{Line}, and the \texttt{Column} \cite{yu2021layout}. \texttt{Group} and \texttt{Line} are horizontal operations, and \texttt{Column} is the vertical operation. \texttt{Group} is a rough characterization (red contours in \figref{fig:characterization}) and the design intuition is that some widgets should be relatively close to each other. Widgets within the same group may occupy different \texttt{Line}s due to their sizes. We further divide the widgets within each \texttt{Group} into different \texttt{Line}s (blue contours in \figref{fig:characterization}) and different \texttt{Column}s (green contours in \figref{fig:characterization}) according to the coordinates. Obtaining \texttt{Column}s in different \texttt{Line}s are conducted respectively, so different \texttt{Line}s may include different amounts of \texttt{Column}s.

\begin{figure}[!htbp]
    \centering
    \includegraphics[width=0.8\linewidth]{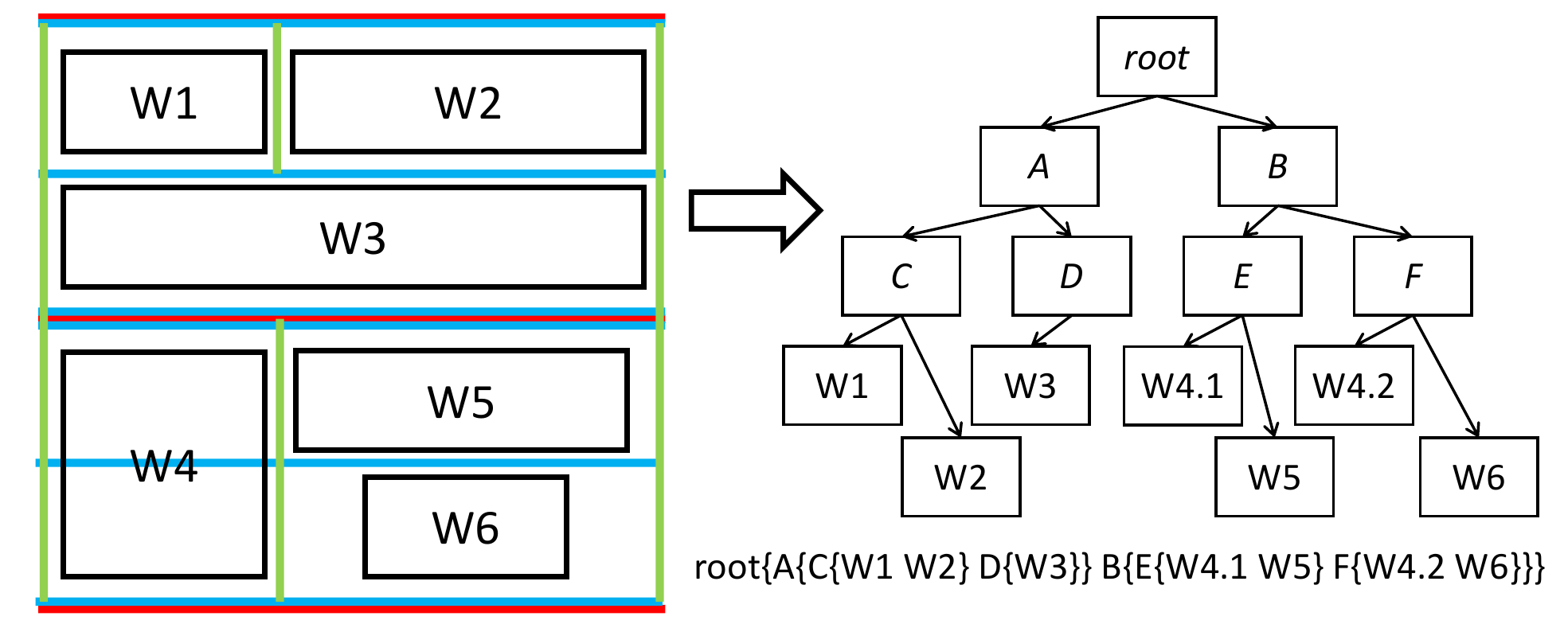}
    \caption{Layout Embedding of App GUI}
    \label{fig:layout}
\end{figure}

The GUI layout views the app GUI from a macroscopic perspective. As described in \secref{sec:layout}, we characterize the layout with a hierarchy structure. The hierarchy structure can be easily transformed into a tree structure as shown in \figref{fig:layout}. The left part of \figref{fig:layout} shows a simplified example of layout characterization, where red lines identify the \texttt{Group}, blue lines identify the \texttt{Line}, and the green lines identify the \texttt{Column}. In the right part of \figref{fig:layout}, the first level represents the whole app screenshot, the second level represents the \texttt{Group}s, the third level represents the \texttt{Line}s, and the fourth level represents the \texttt{Column}s, which are actually the real widgets in the app screenshot. Due to the special design of the layout characterization, the transformed tree structure will have four levels.

To further make use of the layout tree, we transform it into a string. One common practice is to use curly braces to indicate the level relationship. The elements within the curly braces refer to the children nodes of the element before the curly brace. Strings with curly braces can be one-to-one mapped to specific GUI trees.

Then, we embed the GUI layout with the LSTM model \cite{graves2012long}. LSTM is a suitable option because it can effectively embed our input strings indicating the GUI layout structure, which is not quite complex, and it will not bring much extra overhead when processing the GUI layout. LSTM helps achieve a balance between effectiveness and efficiency. In order to train the LSTM model, we randomly generate 100 different GUI layout strings, and the dataset size is 10000 (100 $\times$ 100) pairs. The dataset is divided into the training set, validation set, and test set at the ratio of 7:1:2, following the common practice. The purpose of the LSTM is to fit the tree edit distance \cite{pawlik2015efficient, pawlik2016tree}. The output is a 512-dimension vector after a \texttt{FullyConnected} layer, and the generated vector represents the embedded GUI layout.

In order to study whether there might be potential impact of the random generated strings, we repeat the process for 30 times, following the common practice of software engineering research \cite{arcuri2014hitchhiker}. Since the target of the layout feature extraction model is to fit the tree edit distance, we calculate the sum value of differences between actual distances and predicted distances in each turn. We conduct the t-test on the results of the 30 repetitions, and the null hypothesis is that there is significant difference among different turns. The significance value p is calculated as 0.00018, which is less than 0.05. The null hypothesis is rejected, which means there is no significant difference in the results of different turns.

\subsubsection{\textbf{State Representation}} In order to comprehensively represent the app current state, we combine the microscopic perspective and the macroscopic perspective, \ie the widget embedding and the layout embedding. The widget embedding results and the layout embedding results are merged to obtain the final state (app page) representation. For each widget in the app screenshot, we obtain an 8206-dimension vector. However, the widget amount of different GUI screenshots are different, so if we directly concatenate the vectors, we may get variable-length vectors as widget numbers are different, which do not fit the requirements of the input of the RL framework \cite{romdhana2021deep}. Therefore, we conduct an average calculation of all the widget embedding vectors and obtain one single 8206-dimension vector that represents all the widgets. Then, we concatenate the widget embedding vector with the layout embedding vector to obtain the final app page vector and view it as a corresponding state of the page in the RL framework. The length of the final vector representing the app GUI is 8718 ($2 \times 4096 + 14 + 512$). In this work, we propose a novel app GUI state abstraction algorithm with the GUI image embedding. Our algorithm is designed to simulate the perspective of real app users to view the app GUI, including the widget extraction (microscopic perspective) and the layout characterization (macroscopic perspective). Our work is the first one to combine different granularities of GUI features in to GUI state abstraction, and the results show that the abstraction is effective. Compared with existing learning-based automated testing methods, which directly extract widgets from layout files as the app states, \toolname can better help better characterize the app states from different perspectives, and can more realistically reflect the actual GUI widgets on the app GUI, because some widgets are hidden in other widgets from the layout files (HTML), like the \texttt{Canvas} widgets.

\subsection{Action Generation}
\label{sec:action}

Actions are another significant part of the whole reinforcement learning framework. In the automated software testing scenario, the action is limited. As shown in \tabref{tab:action}, we list all the actions\footnote{``Event'' and ``action'' are equivalent in this paper.} that are considered in \toolname. In software testing, the actions can be divided into three main categories: the widget actions, the page actions, and the system actions. Different actions may have different targets and parameters. In this paper, we consider in total 17 different actions. Some actions apply to both mobile and web platforms, and some are unique to a specific platform.

\begin{table}[!htbp]
\centering
\caption{Actions for \toolname}
\scalebox{1}{
\begin{tabular}{c|c|c|c}
\toprule
Platform & 
Widget Actions & 
Page Actions & 
System Actions \\ \midrule

Both & 
\textit{click}, \textit{input}, \textit{drag}, \textit{double click} & 
\textit{swipe}, \textit{split screen} &
\tabincell{c}{
	\textit{return}, \textit{back switch}, \\ 
	\textit{access grant}, \textit{access deny}, \\ 
	\textit{network switch}
} \\ \midrule 

Mobile & 
\textit{long click} & 
\textit{orientation switch} & 
\textit{phone interrupt} \\ \midrule

Web & 
\textit{mid click}, \textit{right click} & 
\textit{window size} & 
------ 
\\ \bottomrule

\end{tabular}}
\label{tab:action}
\end{table}

Widget actions are actions that are directly applied to specific widgets. We consider the widget image, the widget location, and the widget type presented in \secref{sec:widget} as the target widget embedding feature. Whether a widget action is applicable on the current app page is determined by the widget type (inferred by the CNN model in \secref{sec:widget}). Different types of widget actions can be applied to one widget.

Page actions are actions applied directly to the app GUI page. The page actions mostly involve the page size or orientation changing. For example, the \textit{orientation switch} action will change the app orientation, the \textit{window size} action refers to the window size change. Besides, in order to simplify the action type, we fix some actions. For the \textit{swipe} action, we only consider the up and down swipe directions. Therefore, we set a parameter of page actions as a number that identifies the size change ratio, or the orientations.

System actions refer to the actions that are applied to the operating system. Different from the widget and page actions, system actions are not directly related to the app GUI. No extra information is required to invoke the system actions.

Additionally, the actions are embedded to fit in the Q-network. Based on the analysis of the three categories of actions, action embedding is composed of three parts, the action itself, the target widget, and the parameter information.

For the action itself, we use a 17-dimension one-hot vector to represent the action type. For the target widget, we follow the processing in \secref{sec:widget}, and use an 8206-dimension vector to represent the target widget. For the window size parameter, we take the number parameter as a one-dimension vector to represent it. The three-part vectors are concatenated to obtain the action embedding vector. For some actions, the target widget and the parameter might not be applicable, and in such cases, we will pad the vector with ``0''s in the corresponding places. The action embedding vector is in accordance with the RL framework of \toolname to make the action selection decisions.

For the action design, \toolname considers widget actions, pages actions, and system actions. These actions cover the vast majority of the actions that can be applied to different apps and corresponding GUI widgets. However, most existing learning-based automated testing methods only consider part of the actions, like the ``click'' action. Therefore, we provide a more complete action set to better explore more app states, improving the automated testing efficiency and effectiveness.

\subsection{Reward Calculation}

For each state transition $\langle s_t, a_t, s_{t+1} \rangle$, \toolname assigns a reward. A higher reward means a more significant transition, which may lead to an uncovered state, and then the RL framework learns the state and action values. A well-designed reward function can guide \toolname to explore more state space and avoid being stuck in a local optimum.

At the timestamp $t$, the state transition is $\langle s_t, a_t, s_{t+1} \rangle$, and the \textbf{\textit{transition times}} of $s_t$ is $N_t$, referring to the covered times of the state during the exploration; the executable action number in $s_{t+1}$ is $m_{t+1}$; and the executed action number is $n_{t+1}$. We define the \textbf{\textit{exploration rate}} of state $s_{t+1}$ is $e_{t+1} = \frac{n_{t+1}}{m_{t+1}}$, which means that how much percentage of the actions are executed. Based on the \textbf{\textit{exploration rate}}, the reward is defined as follows:

\begin{equation}
 r_t = \frac{1-e_{t+1}}{\sqrt{N_t}} = \frac{1-\frac{n_{t+1}}{m_{t+1}}}{\sqrt{N_t}}
\end{equation}

Besides, the \textbf{\textit{transition times}} value is added by 1 if the page similarity between the current page and one specific page in exploration memory is confirmed to be higher than a threshold. According to our introduction in \secref{sec:state}, we calculate the page similarity with the widget extraction and layout characterization. With regard to the widget similarity, for all the widgets on page A ($p_A$) and page B ($p_B$), we obtain the one-to-one matched widget pairs if the widgets' Interaction-over-Union (IoU) value \cite{chen2020object} is over the specific threshold (some widgets may not have matching widgets, and such widgets are neglected). For each matched widget pair, we calculate the distance with the embedded widget vector (\secref{sec:widget}) through the Euclidean distance. We define the average distance of all widget pairs as $d(p_A, p_B)$, and the widget similarity is $sim_w(p_A, p_B) = 1 - d(p_A, p_B)$, ranging in [0, 1]. The layout of a GUI page is represented by a string composed of widget nodes and curly braces (\figref{fig:layout}). We use tree edit distance \cite{pawlik2015efficient, pawlik2016tree} to calculate layout similarity between page A $p_A$ and page B $p_B$:

\begin{equation}
sim_l(p_A, p_B) = 1 - \frac{d(A, B)}{max(n_A, n_B)}
\end{equation}

where $d(A, B)$ refers to the tree edit distance, and $n_A$, $n_B$ refer to the node numbers of $p_a$ and $p_b$. Specifically, $max(n_A, n_B)$ actually indicates the theoretical maximum value of $d(A, B)$. $sim_l(p_A, p_B)$ ranges between [0, 1]. Page similarity is the weighted sum of widget and layout similarities. We initially set the 0.5-0.5 weight. We use a threshold to determine whether two pages are the same. In this paper, we use 0.75 as default, following the common practice \cite{yu2021layout}.

In this paper, we design a novel reward function in the RL framework, which comprehensively considers the exploration of both app GUI as a whole and the concrete GUI widgets. The design of the reward function corresponds to our app GUI state abstraction, including the GUI widget (microscopic perspective) and the GUI layout (macroscopic perspective). Compared with existing learning-based automated testing methods, which mostly consider the page similarity only, \toolname can more realistically reflect the app state situation, thus leading to the effective reward assignment and making the exploration effective and efficient.

\subsection{Exploration Policy Execution}
\label{sec:explore}

With the embedded current state and all the applicable actions on existing widgets of the app state, \toolname's exploration policy calculates the values of such actions under such a state, and then executes the appropriate action, with the purpose of exploring as many un-explored states as possible. The exploration policy is a function that gives the execution probability distribution of all possible actions according to the current state. A better designed exploration policy can better guide the exploration of the RL framework on the apps under test.

For automated software testing, the content changes of every accessible app page or widget may cause the app state changes, so the state space to explore is huge. Moreover, the executable action set of different states is different, and the action space of the whole app to explore is huge. Therefore, traditional RL algorithms using the Q-Table will have difficulty processing such a situation. Considering the robust capability of deep learning models, we adopt a deep neural network in \toolname as the key strategy for exploring the app under test, which is supposed to effectively alleviate the explorative space explosion problem.

To determine a test event in the exploration, \toolname first merges the embedded state and each action to form the $\langle state, action \rangle$ pair and then calculates the Q-value of the pair with the Q-network. We add different weights to the Q-values of various kinds of actions in order to reduce the probability of system actions and encourage more exploration within the app itself, and such a practice is widely adopted by literature \cite{zheng2021automatic, pan2020reinforcement}. We set the weight of widget and page actions as 1, and the weight of system actions as 0.5. We set a lower weight for the system actions based on the common practices of existing approaches like Stoat \cite{su2017guided} and Q-testing \cite{pan2020reinforcement}. This practice can avoid executing system actions all the time but not exploring new GUI states. We try different weights from 0.1 to 0.9 at the interval of 0.1 on five apps as a pilot investigation, and we find that 0.5 is about to the optimal value. Therefore, we use the 0.5 as the weight of system actions when compared with the 1 for widget and page actions. With the weighed Q-values obtained, we use the \texttt{SoftMax} function to calculate the execution probability of each action and select which one to execute.

\subsubsection{\textbf{Q-Network}}

The Q-network is implemented with a fully connected neural network with four hidden layers, and each layer has 512 dimensions. The reason why using the fully connected neural network instead of a CNN or RNN model is that the fitting speed is much faster. During the exploration of the RL framework, the Q-network needs to be kept training with constantly refreshed training data, so the fitting speed is important to the efficiency. The fully connected neural network is capable of achieving a balance between high fitting speed and fitting accuracy. The input of the Q-network is the $\langle state, action \rangle$ pair, which is the concatenated result of the embedded state and the embedded action. The output of the Q-network is a real number, indicating the Q-value of the given $\langle state, action \rangle$ pair. Generally speaking, the Q-network takes a vector as the input. The vector is the concatenated state vector and action vector, then the vector goes through four hidden layers, where the parameters are determined by fitting the output Q-values and the expected Q-values of training data. The $\langle state, action \rangle$ pair indicates applying an action on a specific app GUI state. The Q-value indicates the probability to transit to a new app GUI state of the applied action on the specific app GUI state. Then, the output layer is concatenated after the four hidden layers, which outputs a real number, indicating the Q-value of the input $\langle state, action \rangle$ pair. The larger the Q-value is, the higher the probability of exploring new states will be. The loss function of the model is defined as the mean squared error of the output Q-value and the expected Q-value (calculated as \equref{equ:targetq} in \secref{sec:memory}). When exploring the app under test, \toolname constantly re-trains the Q-network to make the fitting more accurate in evaluating the values of executing one specific action. The training process of the Q-network involves the state memory and is illustrated in \secref{sec:memory}. for \toolname, we train the Q-network to fit the Q-values of our novelly designed app GUI states and a more complete action list in the RL framework. We do not use the Q-table because the states of the app require significantly increasing calculation overhead when the enumerated states increase to obtain the state similarity and exploration probability. Considering the features of GUI app state and state transition, especially the huge exploration space, we believe DQN (Q-network) is a suitable choice in the automated GUI testing scenario.

\subsubsection{\textbf{Exploration Memory}}
\label{sec:memory} 

In order to better utilize the explored states to constantly adjust the Q-network, we design the exploration memory to record every state transition and the corresponding reward, which is the data source for the Q-network training. The rewards will have an effect on the exploration policy, and then guide the refreshing of the exploration policy. In the memory, the following data of timestamp $t$ are recorded: the state transition $\langle s_t, a_t, s_{t+1} \rangle$, the reward $r_t$, and the action set $A_{t+1}$ of state $s_{t+1}$. $s_t$, $a_t$, and $r_t$ represent the state embedding, action embedding, and the reward at the timestamp $t$, respectively. $A_{t+1}$ is the executable action set at the timestamp $t+1$. Such information can be used to calculate the $Q^{target}$, which is the expected Q-value and can guide the fitting of the Q-network (\equref{equ:targetq}).

\begin{equation}
  \label{equ:targetq}
  Q_{t}^{target} = r_{t} + \gamma\text{max}_{a \in A_{t+1}} Q\left(s_{t+1}, a\right)
\end{equation}

Parameter $\gamma$ is the decay rate, and a larger $\gamma$ means that a reward of a more recent reward is more important. With the further exploration of \toolname, more and more data are recorded. Considering the efficiency problem, we cannot take all the data into the Q-network training. We have to select part of the data as the training data. In order to instruct the Q-network to tendentiously learn from recent state transitions, we use the most recent $\frac{n}{2}$ data and randomly select another $\frac{n}{2}$ data from the rest of data.

\subsubsection{\textbf{Randomization Strategy}}

\toolname can get the execution probability of all the actions under the current state with the values of the $\langle state, action \rangle$ pairs, and it is optimal to execute the action with the highest Q-value. However, if the optimal actions are always executed, the framework keeps exploring with existing knowledge, making it hard to explore unknown states. This is the so-called ``exploration-exploitation dilemma'' \cite{robbins1952some} in RL. For \toolname, we use the Boltzmann strategy \cite{cesa2017boltzmann} to alleviate the problem. The Boltzmann strategy assigns a probability to each $\langle state, action \rangle$ pair according to the Q-value. Then, the probabilities will directly affect which action is to be executed.

\subsection{Implementation}

In order to devote the proposed platform-independent reinforcement learning framework into practice, we design and implement the corresponding tool. In this section, we illustrate the concrete design and detailed algorithms and parameters. For the app GUI state abstraction, we use the VGG-16 model to extract the features. We follow the default training settings of hyper-parameters. The batch size is 256, the training epoch is 5, and the learning rate is 0.001. The training is conducted on a Ubuntu 20.04.2 machine with the GTX 1080Ti GPU. For the Q-network, we set the batch size as 64, the training epoch as 5, and the learning rate as 0.01, following the existing studies like \qtesting \cite{pan2020reinforcement}.

\section{Experiment}

\subsection{Experiment Setting}

\subsubsection{\textbf{Research Question}}

We set three research questions to evaluate the \toolname, and the first two questions focus on the code coverage, bug detection capability, and corresponding cross analyses on mobile apps and web apps, respectively. The third question has an in-depth insight into the advantages of \toolname. 

\begin{itemize}
  \item \textbf{RQ1 (Code Coverage)}: How effectively can \toolname cover app source code on different apps?
  \item \textbf{RQ2 (Bug Detection)}: How effectively can \toolname detect real-world bugs on different apps? 
  \item \textbf{RQ3 (Advantage Analysis)}: How can \toolname outperform the baselines?
\end{itemize}

\subsubsection{\textbf{Experiment Subjects}}

For mobile apps, we collect real-world apps from existing studies, including \cite{mao2016sapienz, su2017guided, pan2020reinforcement}. Besides, We use widely-used open-source apps from a popular list \cite{liu2020androzooopen}. For web apps, we use the apps in existing work like \cite{zheng2021automatic}. However, due to the maintenance problem of the apps, some apps cannot be compiled or instrumented, so we have to only use the usable ones. For the mobile apps, we first use a Samsung mobile device with Android 11, some apps are out-of-date, no longer maintained to be executed, and are not supported on this Android version, so we remove such apps. Then, due to the requirement to obtain code coverage and running logs, we use a widely used tool to intrude the apps, but some apps cannot be instrumented to obtain the code coverage, so we also remove such apps. Finally, we use 20 mobile apps in our evaluation. For the web apps, we consider the full source code, including front end and server end, to obtain the code coverage information, which is not used as the evaluation metric in existing studies, which only considers the HTML page coverage, so we choose the source-code-available apps in our evaluation. We respectively instrument the mobile apps with Jacoco\footnote{https://www.jacoco.org/jacoco/} and web apps with NCY\footnote{https://istanbul.js.org/}. More details of our experiment subject apps are available on our online package due to the page limit. 

\subsubsection{\textbf{Baseline Approaches}}

In order to evaluate the capability of \toolname more comprehensively, we select 9 baselines for mobile app GUI testing and 2 baselines for web app GUI testing, including random-based ones, model-based ones, and learning-based ones. We use the \monkey \cite{google2022monkey} as a representative of random strategy. \monkey is originally used for mobile apps, so we implement the \monkey for the web platform. For the model-based strategy for mobile apps, we use four representative tools, Stoat \cite{su2017guided}, TimeMachine \cite{dong2020time}, ComboDroid \cite{wang2020combodroid}, and Ape \cite{gu2019practical}, as the baselines. These baselines are representative and state-of-the-art tools for mobile app exploration testing. For the learning-based strategy, we use four state-of-the-art approaches, the Humanoid \cite{li2019humanoid}, Q-testing \cite{pan2020reinforcement}, GPTDroid \cite{liu2023chatting}, and the WebExplor \cite{zheng2021automatic}, which are cutting-edge technologies for automated mobile and web app testing. We believe such tools, which have proven their capability compared with other tools, are representative. With regard to the tool implementation, for \monkey on the mobile platform, the tool is directly usable provided by Google. For the \monkey on the web platform, we implement a tool imitating the strategy adopted by Google \monkey, guaranteeing that it can randomly generate test events for web app exploration.  Some baselines, like the Stoat and Q-testing, are available because the authors open-source them, while some others are not available, like the GPTDroid, so we have to replicate them by ourselves, and we have tried our best to follow the paper descriptions to reproduce the tool. Experiments are repeated 5 times and we apply the Wilcoxon non-parametric statistical test with the consideration of the non-determinism of the algorithms. The average results are recorded when the p-value is less than 0.05.

\subsubsection{\textbf{Evaluation Metrics}}

Basically, we adopt the most widely accepted metric, coverage, as the main metric of this paper. Specifically, we utilize the coverage of two granularities, the line coverage and the branch coverage. We use the results of running 1.5 hours, 2 hours, 3 hours, and generate 500 test events, following the common practice like in \cite{pan2020reinforcement, su2017guided, zheng2021automatic}. One thing to notice is that if less than 500 events are generated within 1.5 hours, we will continue until 500 events are generated. During the comparison of 1.5-hour, 2-hour, and 3-hour results, we do not count how many events are generated. We evaluate the bug detection capability of \toolname on real-world apps by calculating how many bugs \toolname can detect. We identify crashes that are actually caused by the app under test by monitoring Logcat information, and we conduct the deduplication by analyzing the information regarding app activity, app package, and stack data. Further, we analyze the cross situation of code coverage and detected bugs, which means how many code lines/branches and detected bugs are covered by both tools or how many code lines/branches and detected bugs are additionally covered by one tool than that of another one. Specifically, cross bug detection is calculated according to the bug quality. The cross code coverage is denoted as A $\uparrow$ B, which means the percentage of  additional covered code of a tool to the all covered code of such a tool. A $\uparrow$ B is calculated as follows: 

\begin{equation}
  A \uparrow B = \frac{\text{coverage of tool A} - \text{coverage of tool B}}{\text{coverage of tool A}} \times 100\%
\end{equation}

\begin{table}[!htbp]
\centering
\caption{Summary of Code Coverage \& Bug Detection Capability Comparison: Average Code Coverage and Detected Bug Summation (M: Monkey, S: Stoat, T: TimeMachine, C: ComboDroid, H: Humanoid, A: APE, Q: Q-testing, G: GPTDroid, W: WebExplor, P: \toolname)}
\scalebox{0.75}{
\begin{tabular}{c|ccccccccc|ccc}

\toprule
\multirow{2}{*}{Configuration} & 
\multicolumn{9}{c|}{Mobile App} & 
\multicolumn{3}{c}{Web App} \\ \cmidrule{2-13}
 & M & S & T & C & H & A & Q & G & P & M & W & P \\ \midrule
Line Coverage (500step)   & 23.7\% & 23.7\% & 24.2\% & 24.7\% & 25.4\% & 21.9\% & 24.0\% & 27.9\% & 30.8\% & 21.5\% & 30.6\% & 31.0\% \\
Line Coverage (1.5h)      & 25.6\% & 25.5\% & 24.7\% & 24.9\% & 25.3\% & 26.1\% & 30.7\% & 32.4\% & 34.7\% & 28.3\% & 25.5\% & 29.4\% \\
Line Coverage (2h)        & 34.9\% & 35.4\% & 35.0\% & 35.5\% & 36.1\% & 35.5\% & 40.6\% & 44.7\% & 49.8\% & 37.7\% & 33.6\% & 37.8\% \\
Line Coverage (3h)        & 35.4\% & 35.9\% & 35.4\% & 36.0\% & 36.6\% & 36.0\% & 41.0\% & 45.2\% & 50.4\% & 42.9\% & 41.8\% & 48.8\% \\ \midrule
Branch Coverage (500step) & 12.0\% & 12.5\% & 13.1\% & 13.5\% & 14.7\% & 14.7\% & 12.4\% & 16.2\% & 16.9\% & 12.7\% & 18.7\% & 19.2\% \\
Branch Coverage (1.5h)    & 14.0\% & 14.8\% & 15.5\% & 15.7\% & 15.8\% & 16.5\% & 18.6\% & 18.9\% & 19.7\% & 12.6\% & 10.8\% & 13.2\% \\
Branch Coverage (2h)      & 19.6\% & 20.2\% & 21.2\% & 20.9\% & 20.0\% & 22.1\% & 22.9\% & 23.1\% & 24.1\% & 18.1\% & 16.1\% & 18.6\% \\
Branch Coverage (3h)      & 22.9\% & 24.2\% & 24.4\% & 23.8\% & 24.3\% & 26.4\% & 26.8\% & 25.7\% & 28.9\% & 19.4\% & 17.0\% & 19.7\% \\ \midrule
Detected Bug (500step) & 20 & 52 & 52 & 50 & 43 & 44 & 49  & 97  & 117 & 1 & 2 & 6 \\
Detected Bug (1.5h)    & 34 & 56 & 58 & 57 & 55 & 49 & 116 & 96  & 121 & 1 & 2 & 4 \\
Detected Bug (2h)      & 45 & 65 & 68 & 67 & 65 & 56 & 127 & 105 & 132 & 4 & 5 & 11 \\
Detected Bug (3h)      & 62 & 78 & 75 & 76 & 74 & 68 & 136 & 113 & 145 & 4 & 8 & 13 \\ \midrule
\end{tabular}}
\label{tab:expsum}
\end{table}

\subsection{RQ1: Code Coverage}

\begin{figure}[!htbp]
	\subfigure[Line Coverage (500step)]{\begin{minipage}[t]{0.23\linewidth}\centering\includegraphics[width=\linewidth]{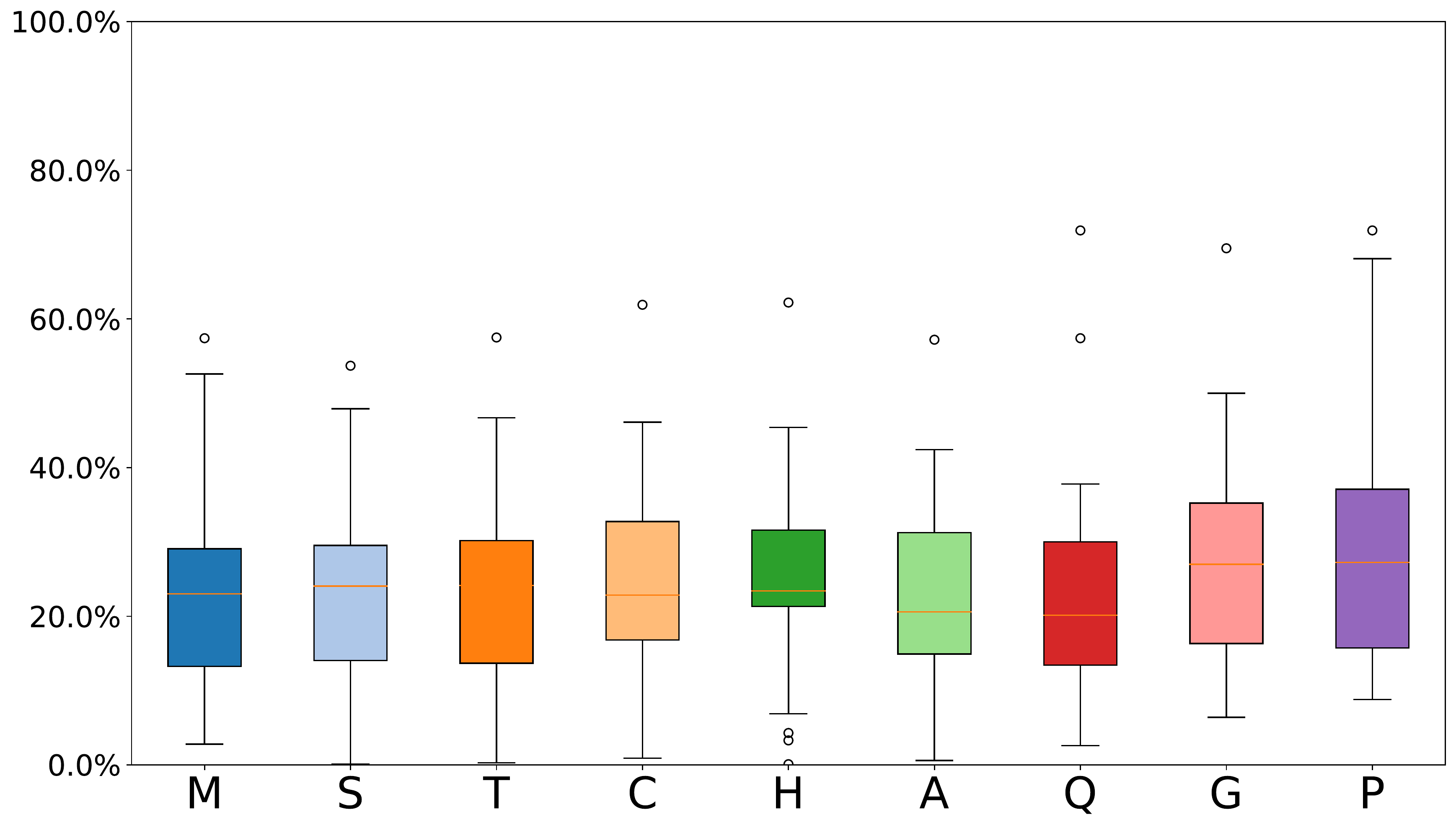}\end{minipage}}
	\hfill
	\subfigure[Line Coverage (1.5h)]{\begin{minipage}[t]{0.23\linewidth}\centering\includegraphics[width=\linewidth]{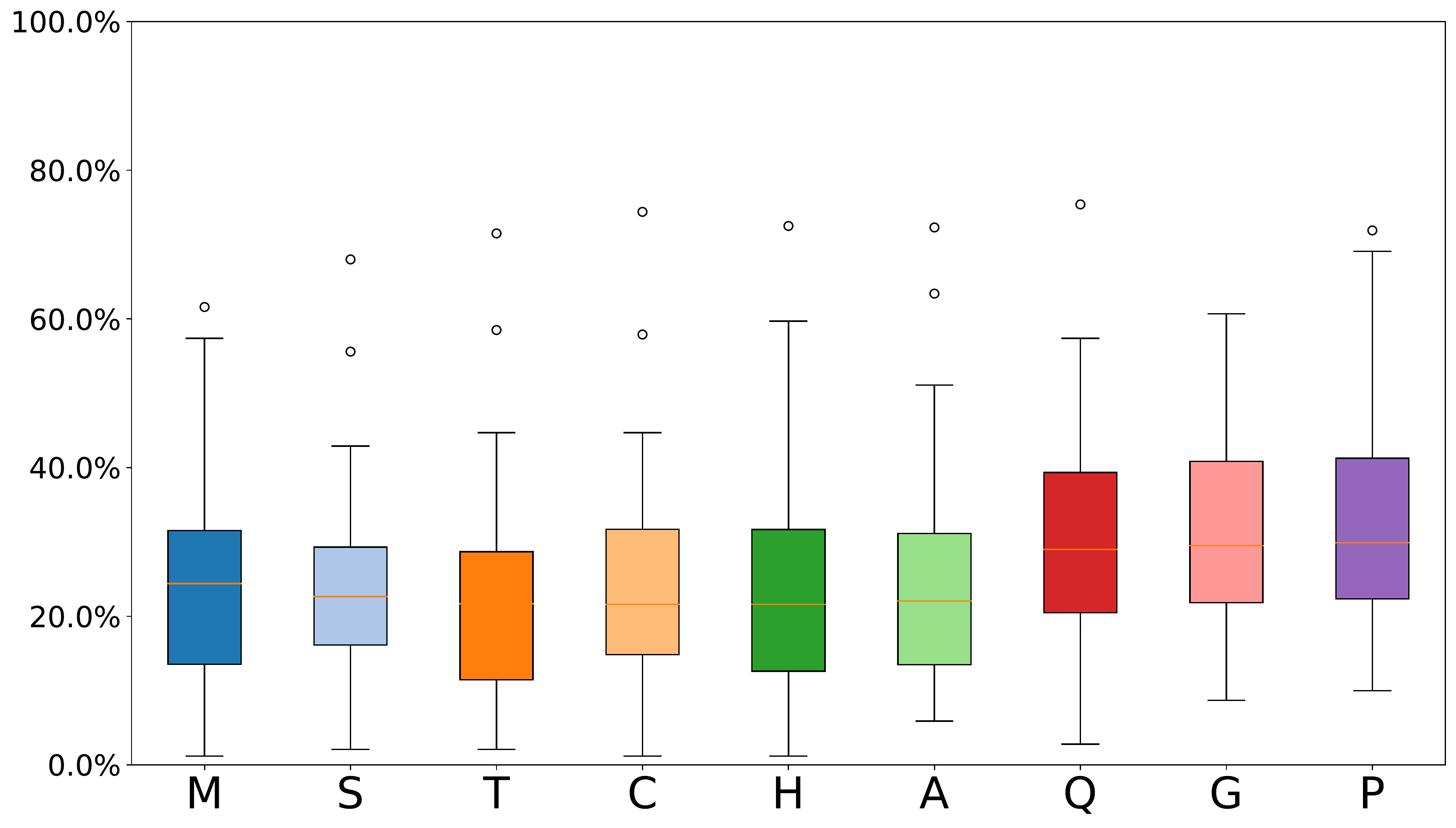}\end{minipage}}
	\hfill
	\subfigure[Line Coverage (2h)]{\begin{minipage}[t]{0.23\linewidth}\centering\includegraphics[width=\linewidth]{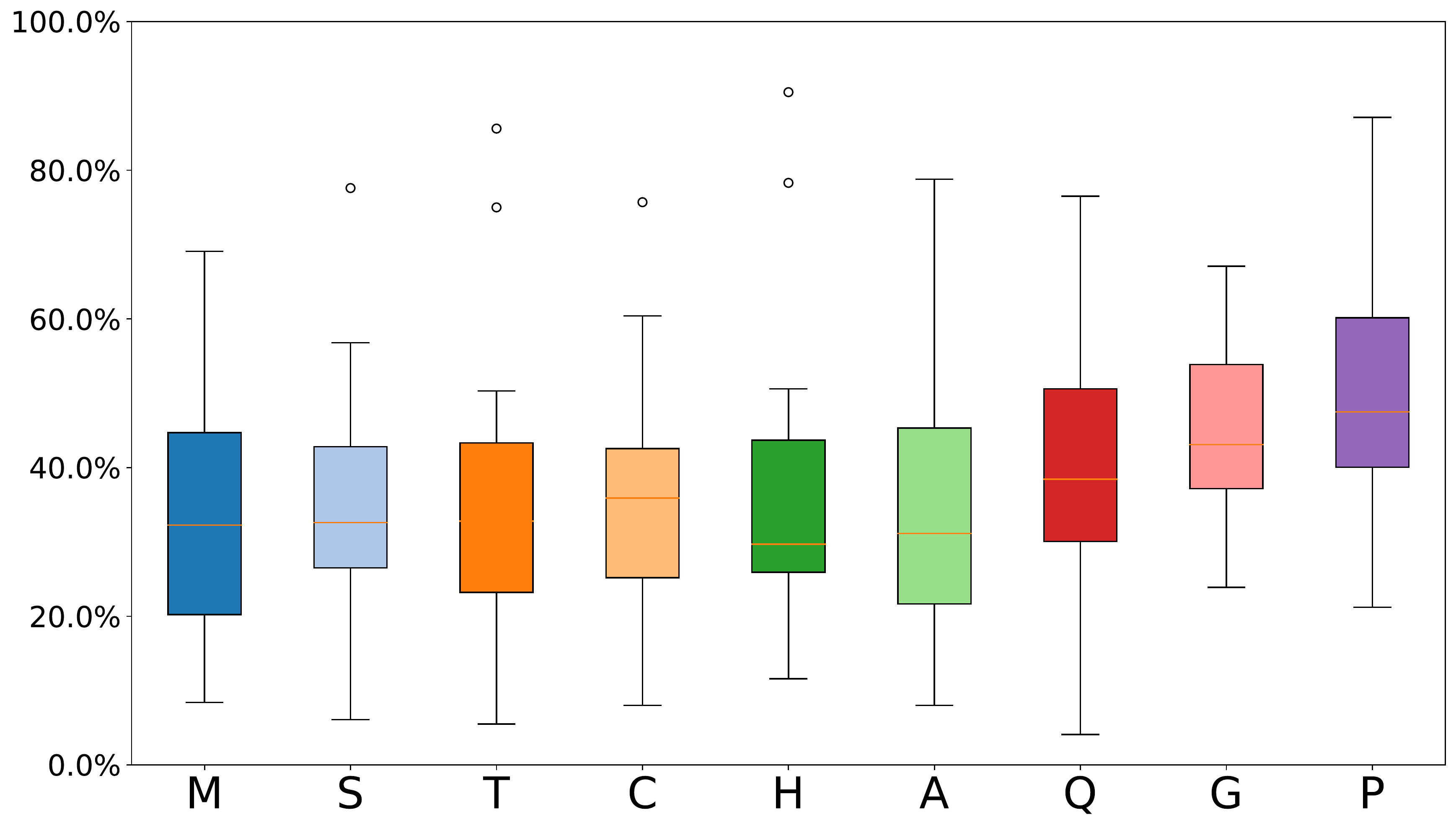}\end{minipage}}
	\hfill
	\subfigure[Line Coverage (3h)]{\begin{minipage}[t]{0.23\linewidth}\centering\includegraphics[width=\linewidth]{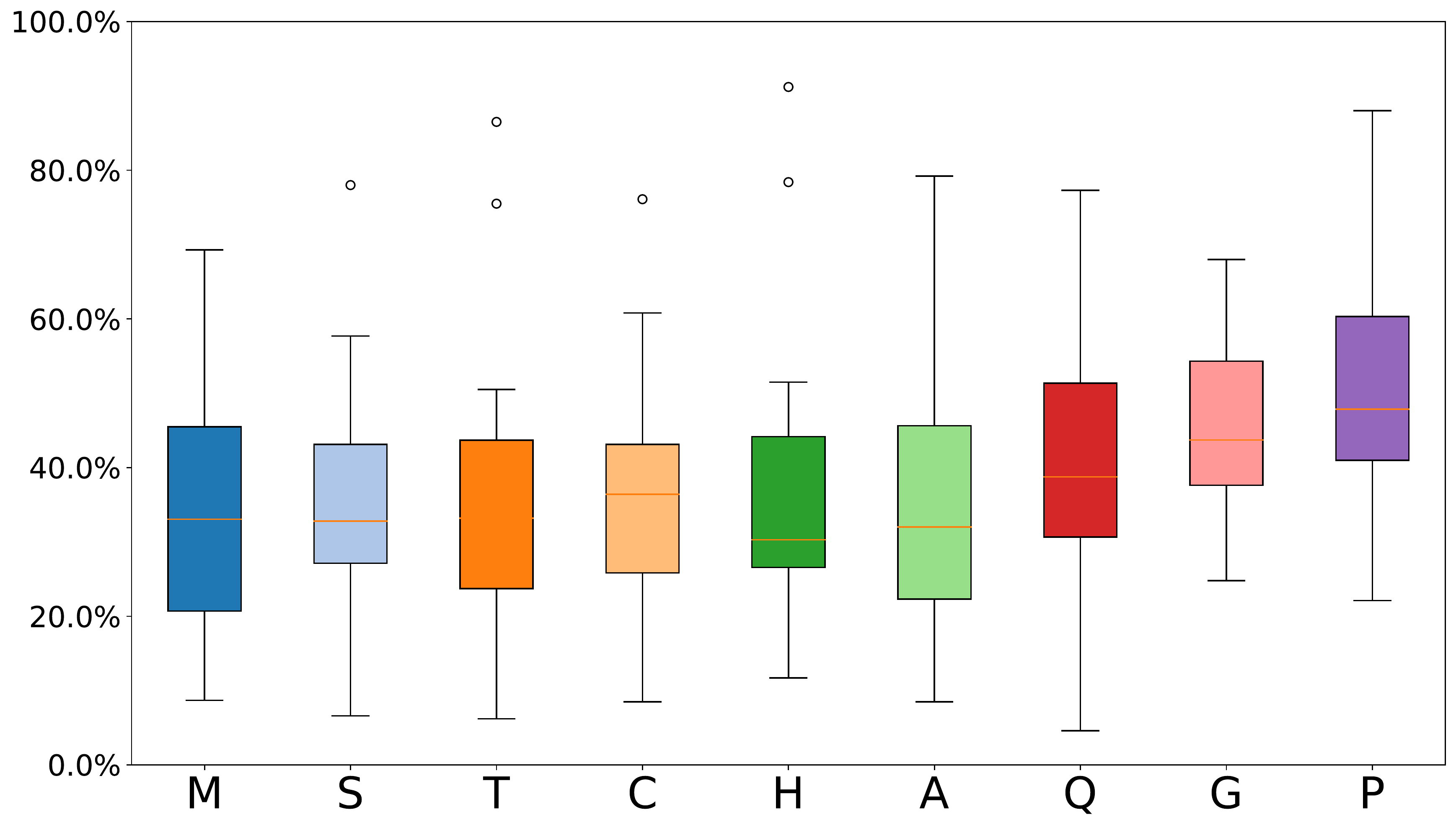}\end{minipage}}
	
	\subfigure[Branch Coverage (500step)]{\begin{minipage}[t]{0.23\linewidth}\centering\includegraphics[width=\linewidth]{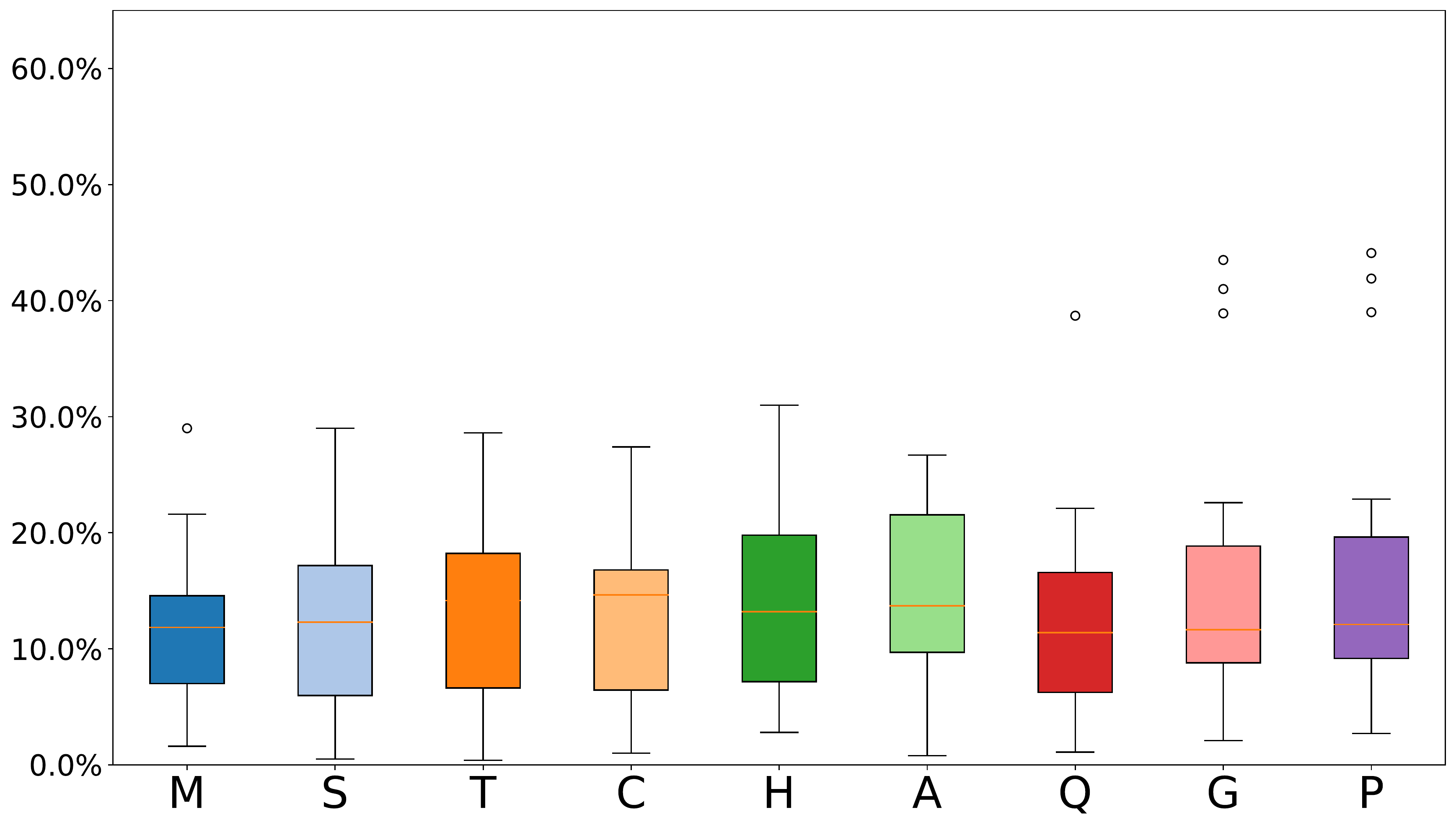}\end{minipage}}
	\hfill
	\subfigure[Branch Coverage (1.5h)]{\begin{minipage}[t]{0.23\linewidth}\centering\includegraphics[width=\linewidth]{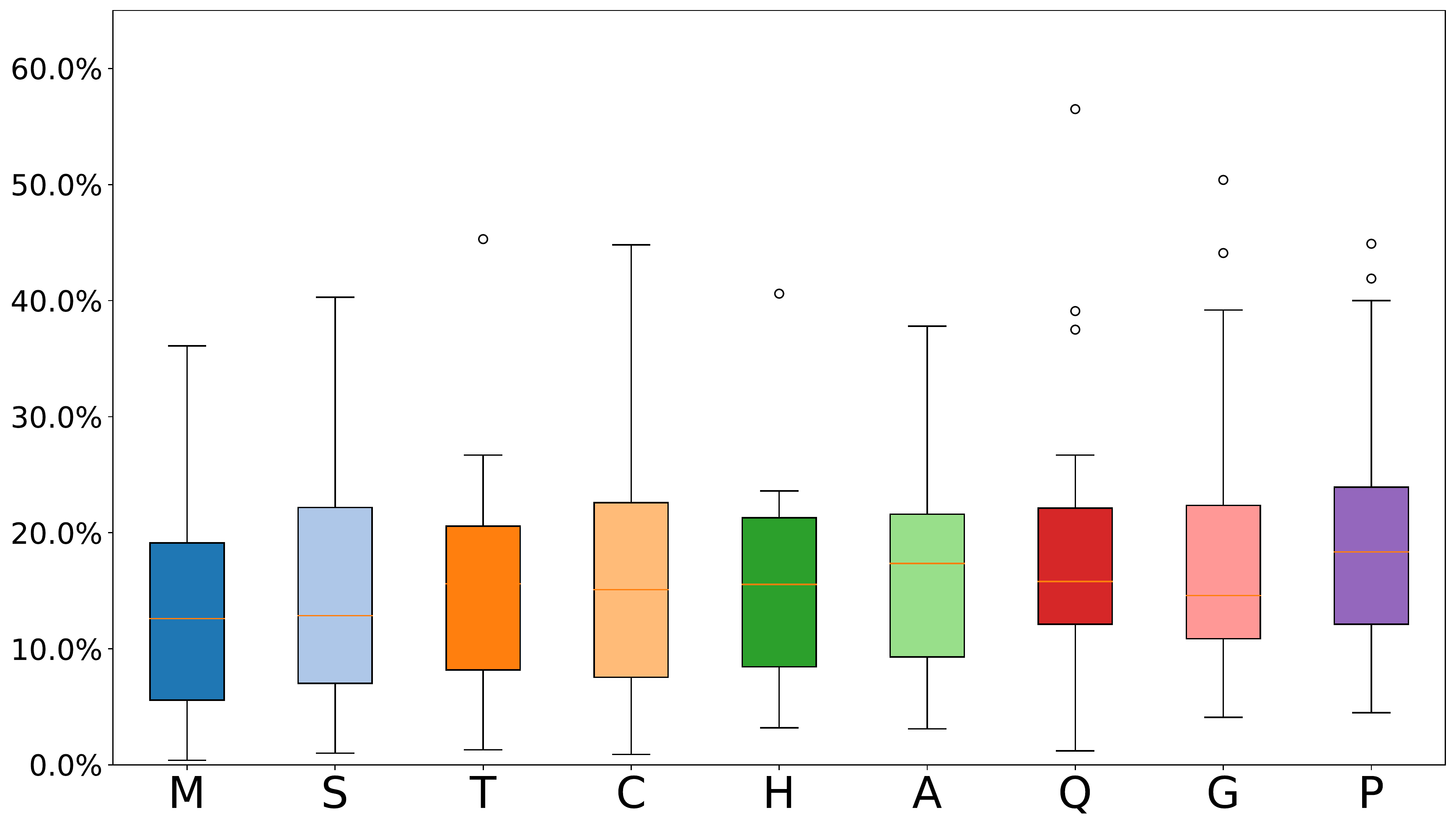}\end{minipage}}
	\hfill
	\subfigure[Branch Coverage (2h)]{\begin{minipage}[t]{0.23\linewidth}\centering\includegraphics[width=\linewidth]{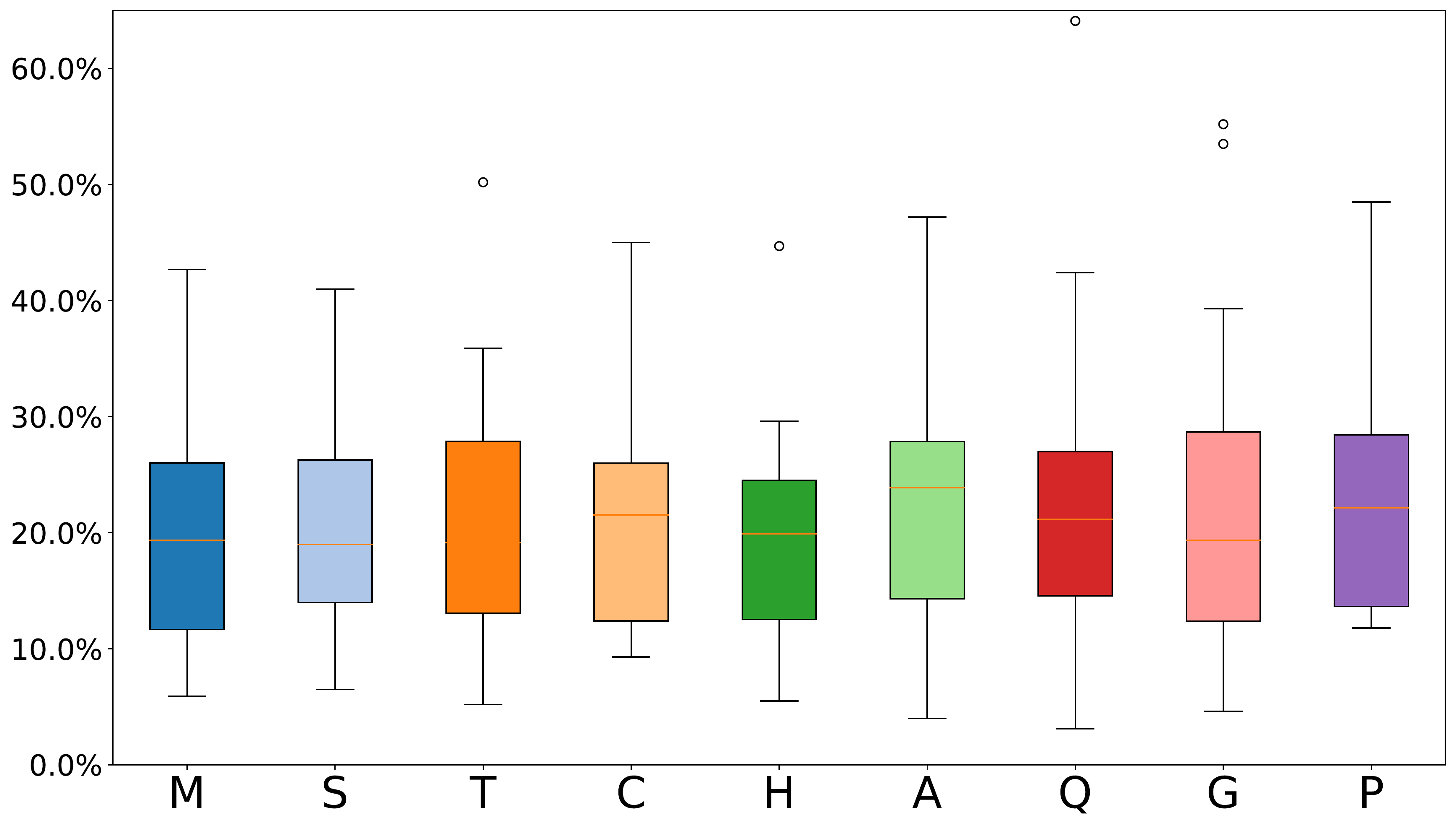}\end{minipage}}
	\hfill
	\subfigure[Branch Coverage (3h)]{\begin{minipage}[t]{0.23\linewidth}\centering\includegraphics[width=\linewidth]{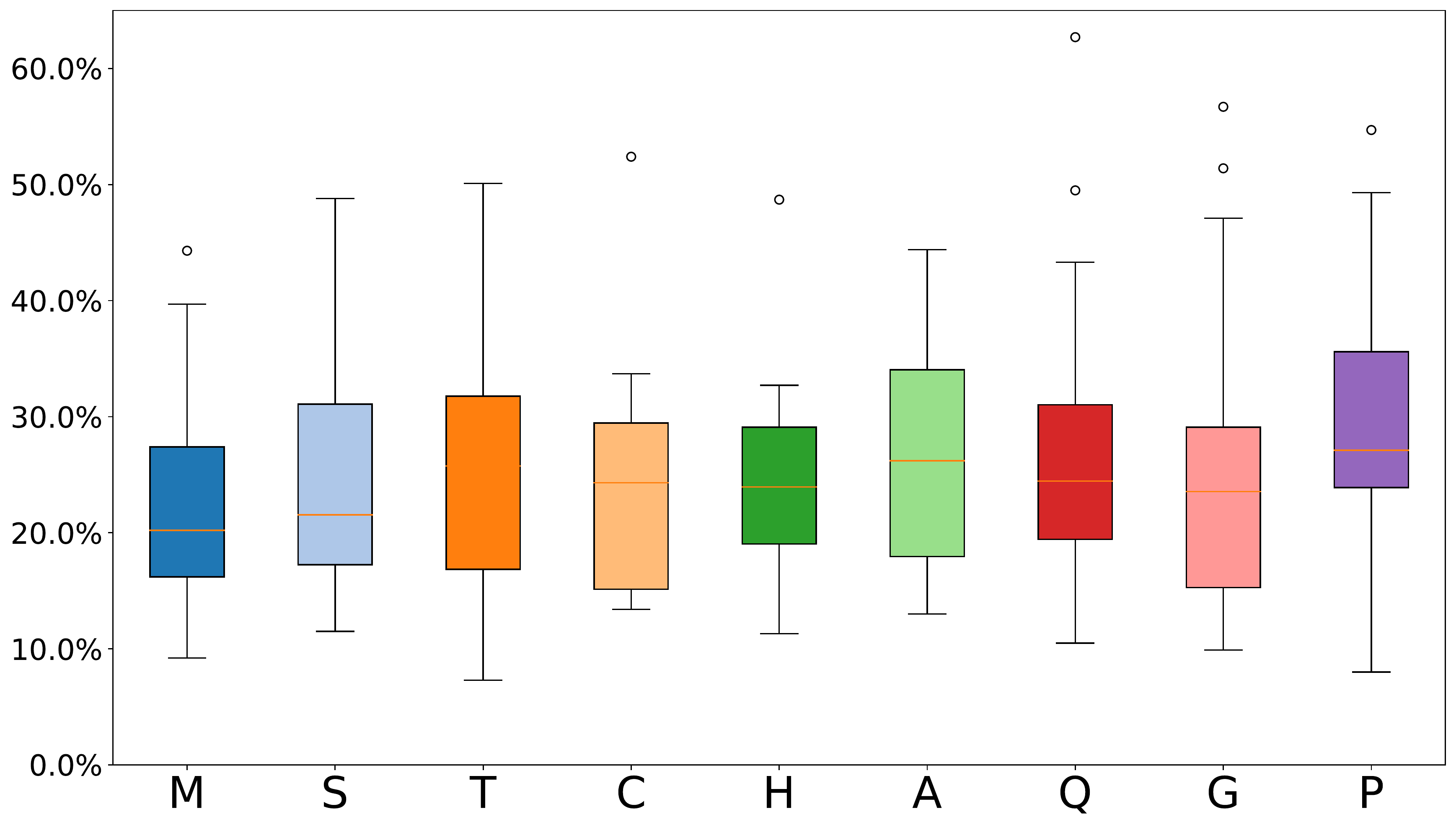}\end{minipage}}
	
	\subfigure[Detected Bug (500step)]{\begin{minipage}[t]{0.23\linewidth}\centering\includegraphics[width=\linewidth]{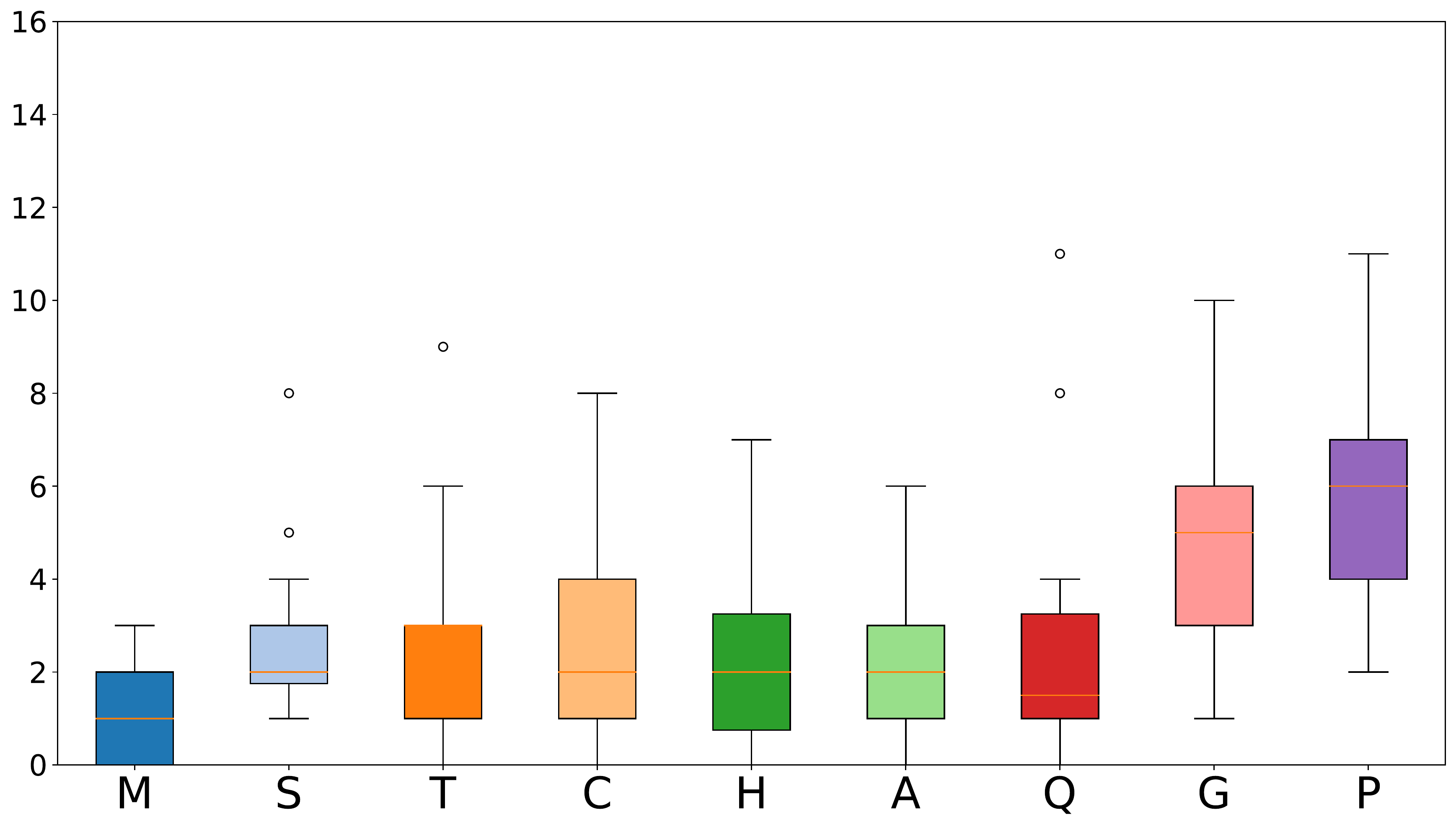}\end{minipage}}
	\hfill
	\subfigure[Detected Bug (1.5h)]{\begin{minipage}[t]{0.23\linewidth}\centering\includegraphics[width=\linewidth]{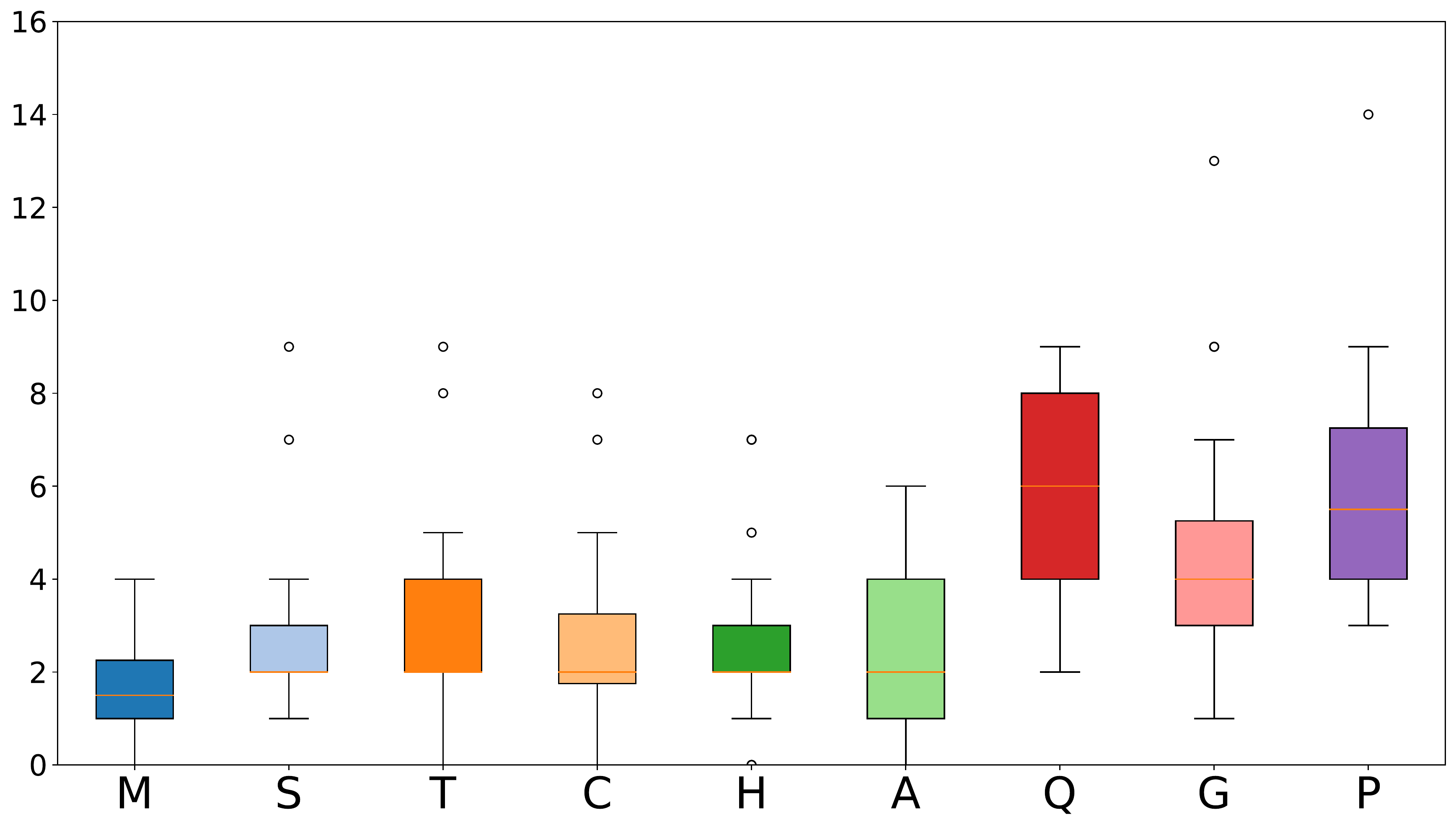}\end{minipage}}
	\hfill
	\subfigure[Detected Bug (2h)]{\begin{minipage}[t]{0.23\linewidth}\centering\includegraphics[width=\linewidth]{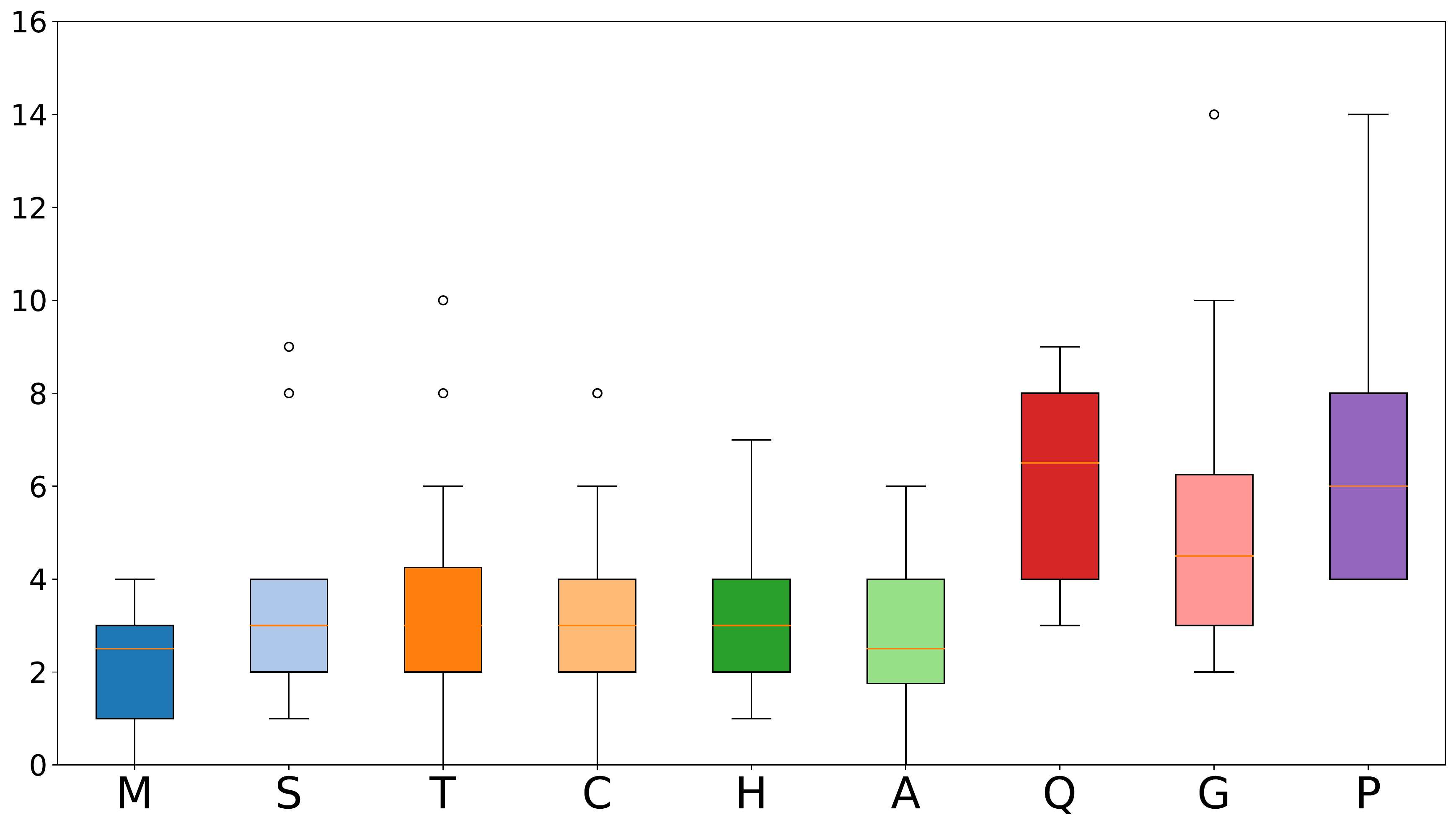}\end{minipage}}
	\hfill
	\subfigure[Detected Bug (3h)]{\begin{minipage}[t]{0.23\linewidth}\centering\includegraphics[width=\linewidth]{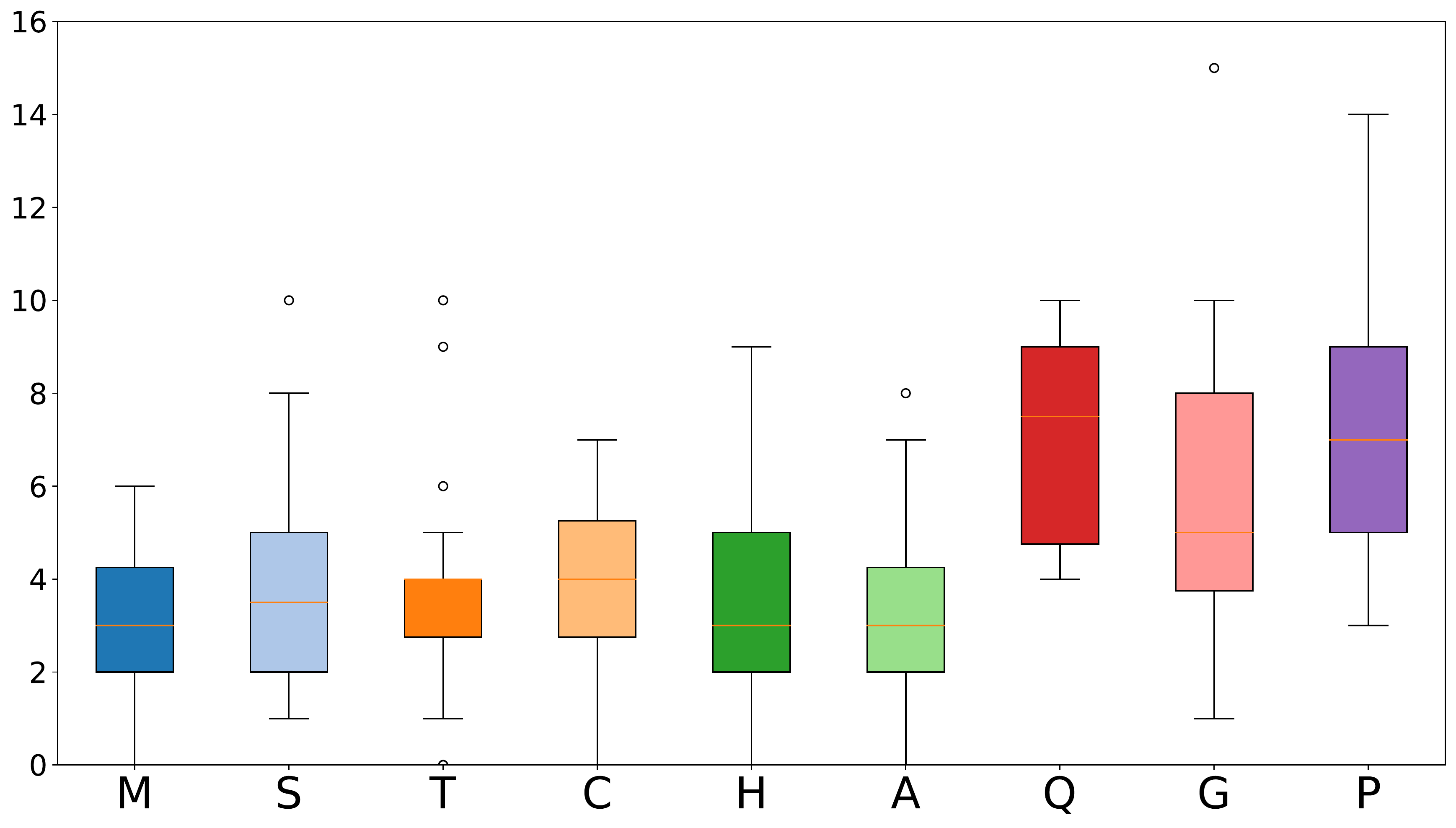}\end{minipage}}
	\centering
	\caption{Code Coverage \& Bug Detection Capability Comparison on Mobile Apps (M: Monkey, S: Stoat, T: TimeMachine, C: ComboDroid, H: Humanoid, A: APE, Q: Q-testing, G: GPTDroid, P: \toolname)}
\label{fig:expmob}
\end{figure}

\begin{figure}[!htbp]
	\subfigure[Line Coverage (500step)]{\begin{minipage}[t]{0.23\linewidth}\centering\includegraphics[width=\linewidth]{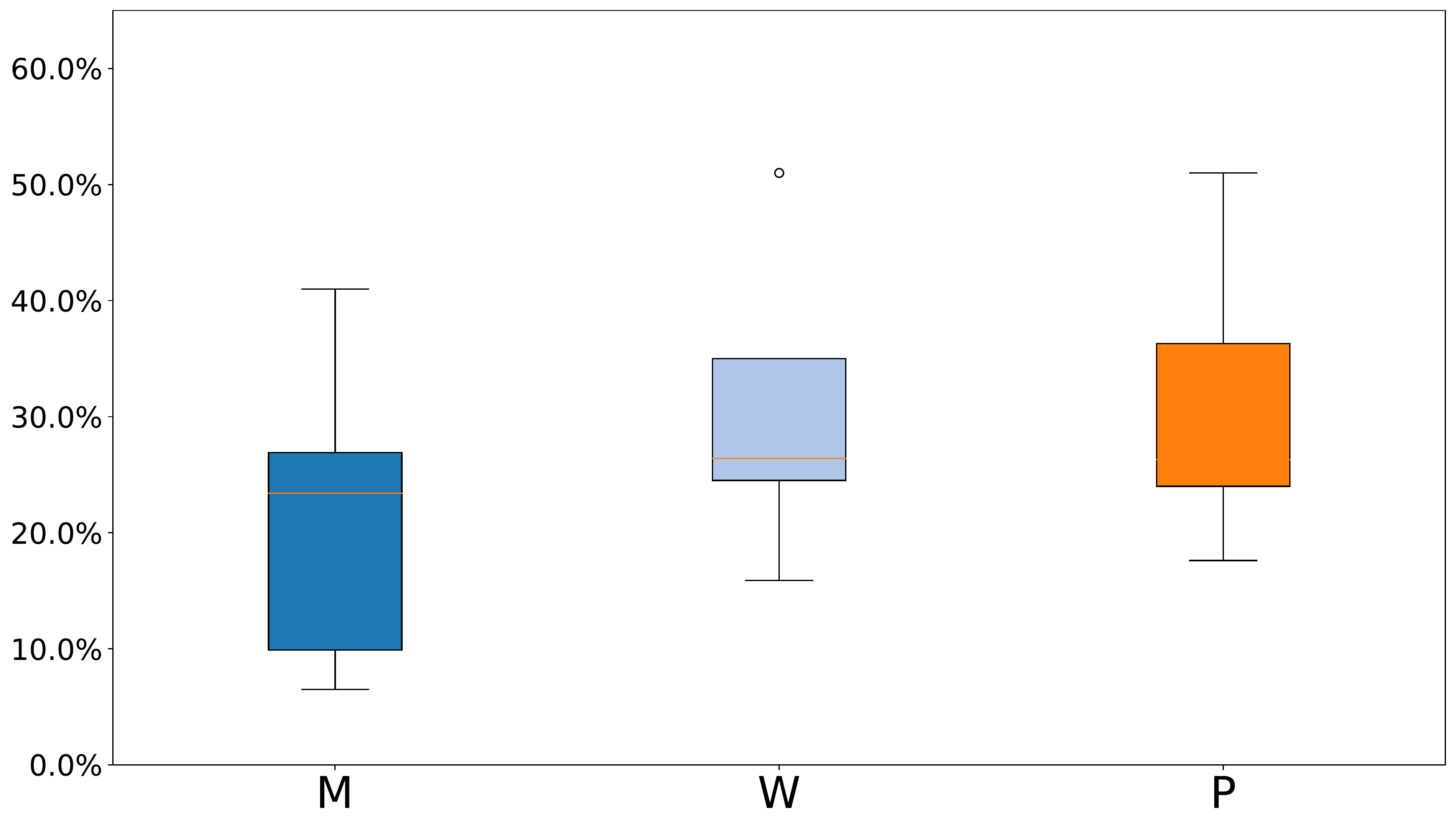}\end{minipage}}
	\hfill
	\subfigure[Line Coverage (1.5h)]{\begin{minipage}[t]{0.23\linewidth}\centering\includegraphics[width=\linewidth]{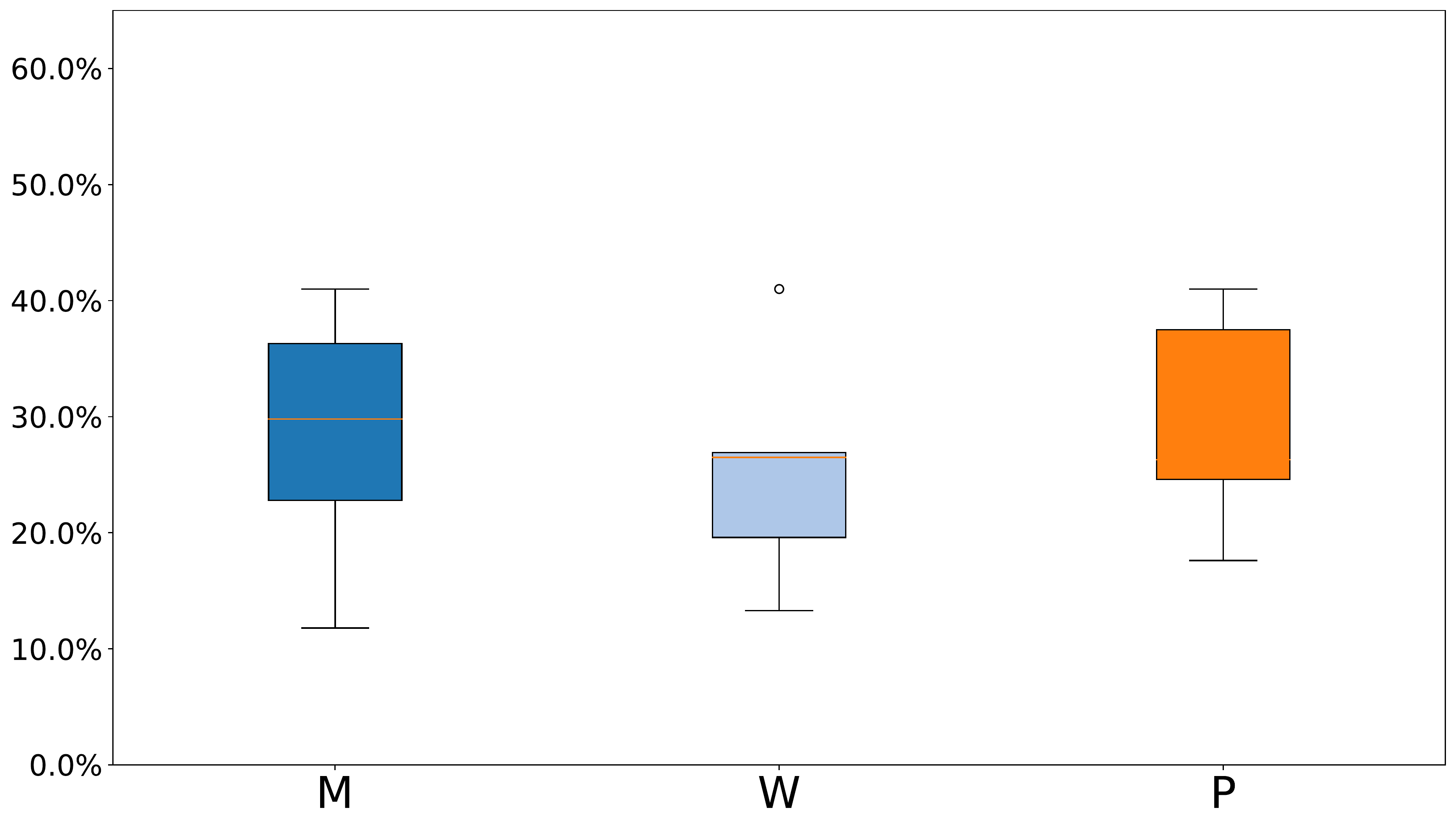}\end{minipage}}
	\hfill
	\subfigure[Line Coverage (2h)]{\begin{minipage}[t]{0.23\linewidth}\centering\includegraphics[width=\linewidth]{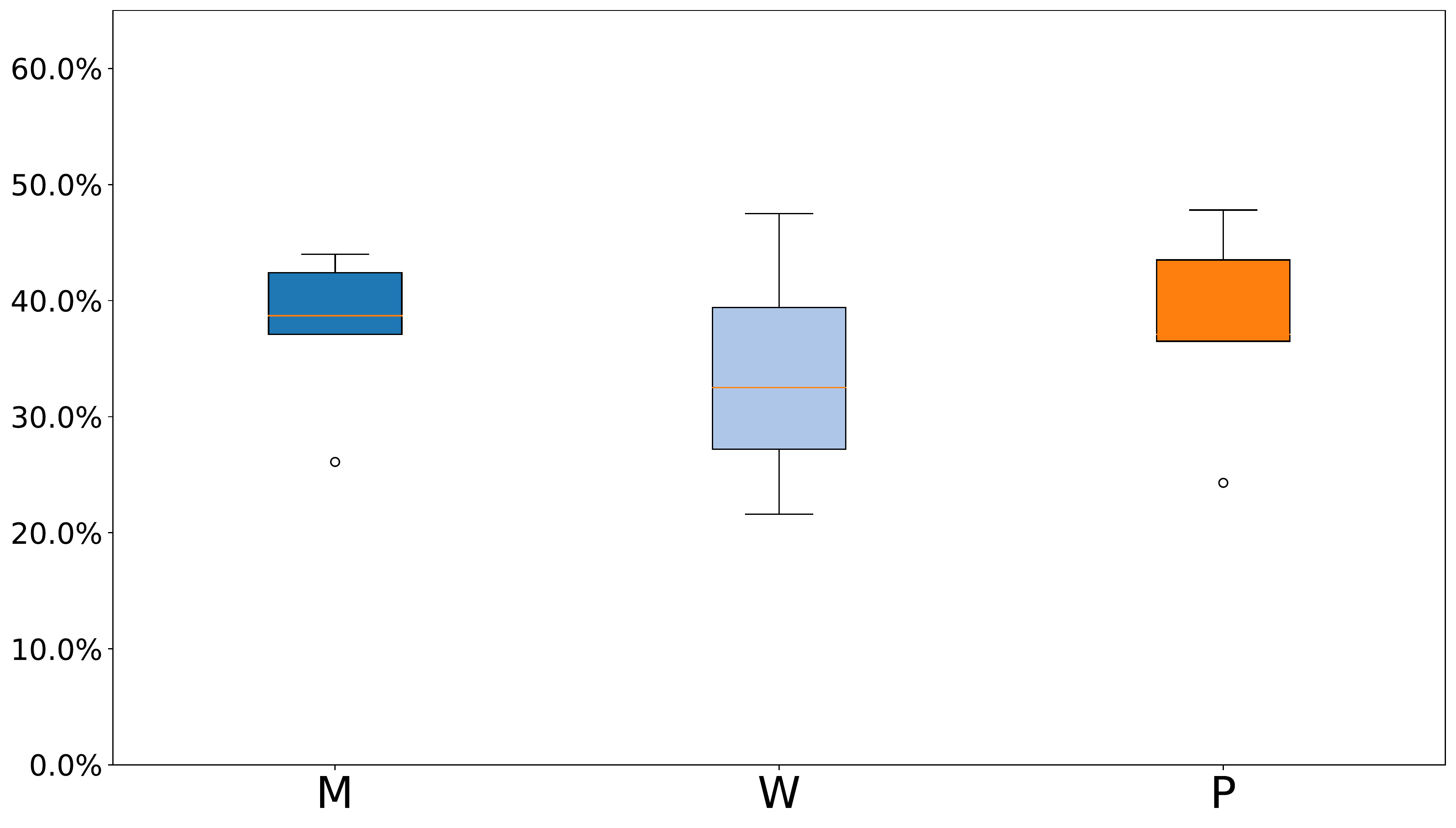}\end{minipage}}
	\hfill
	\subfigure[Line Coverage (3h)]{\begin{minipage}[t]{0.23\linewidth}\centering\includegraphics[width=\linewidth]{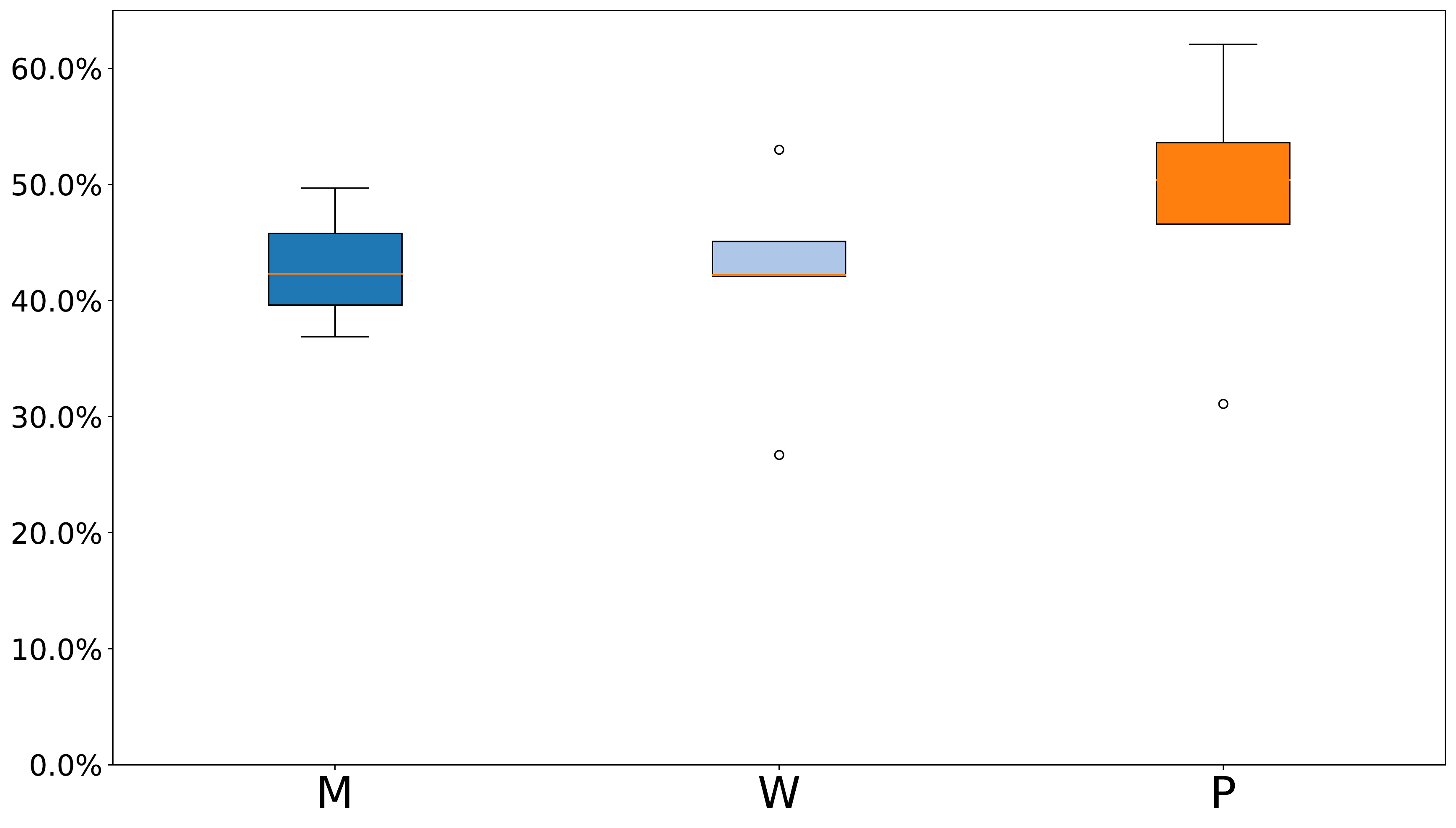}\end{minipage}}
	
	\subfigure[Branch Coverage (500step)]{\begin{minipage}[t]{0.23\linewidth}\centering\includegraphics[width=\linewidth]{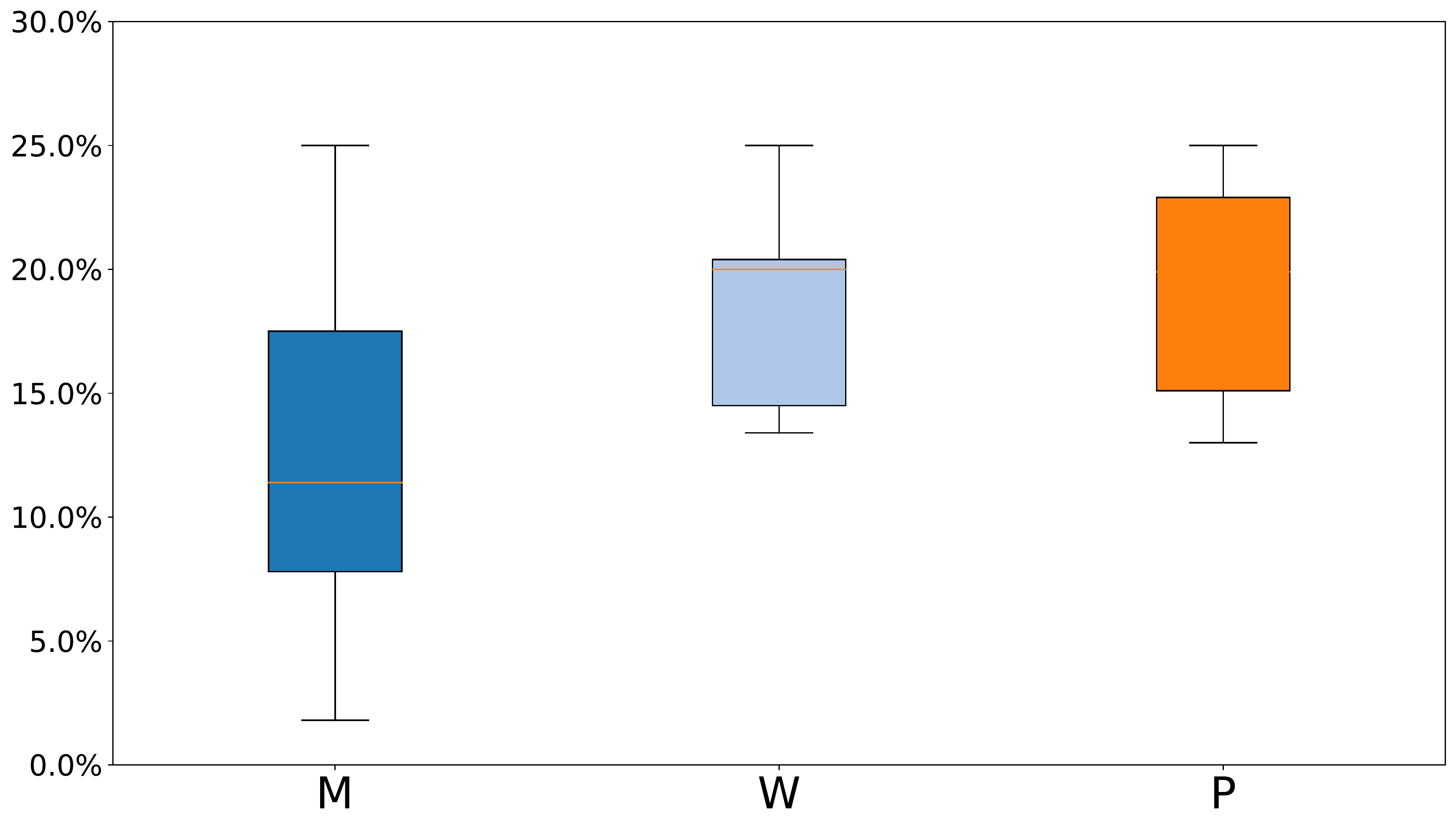}\end{minipage}}
	\hfill
	\subfigure[Branch Coverage (1.5h)]{\begin{minipage}[t]{0.23\linewidth}\centering\includegraphics[width=\linewidth]{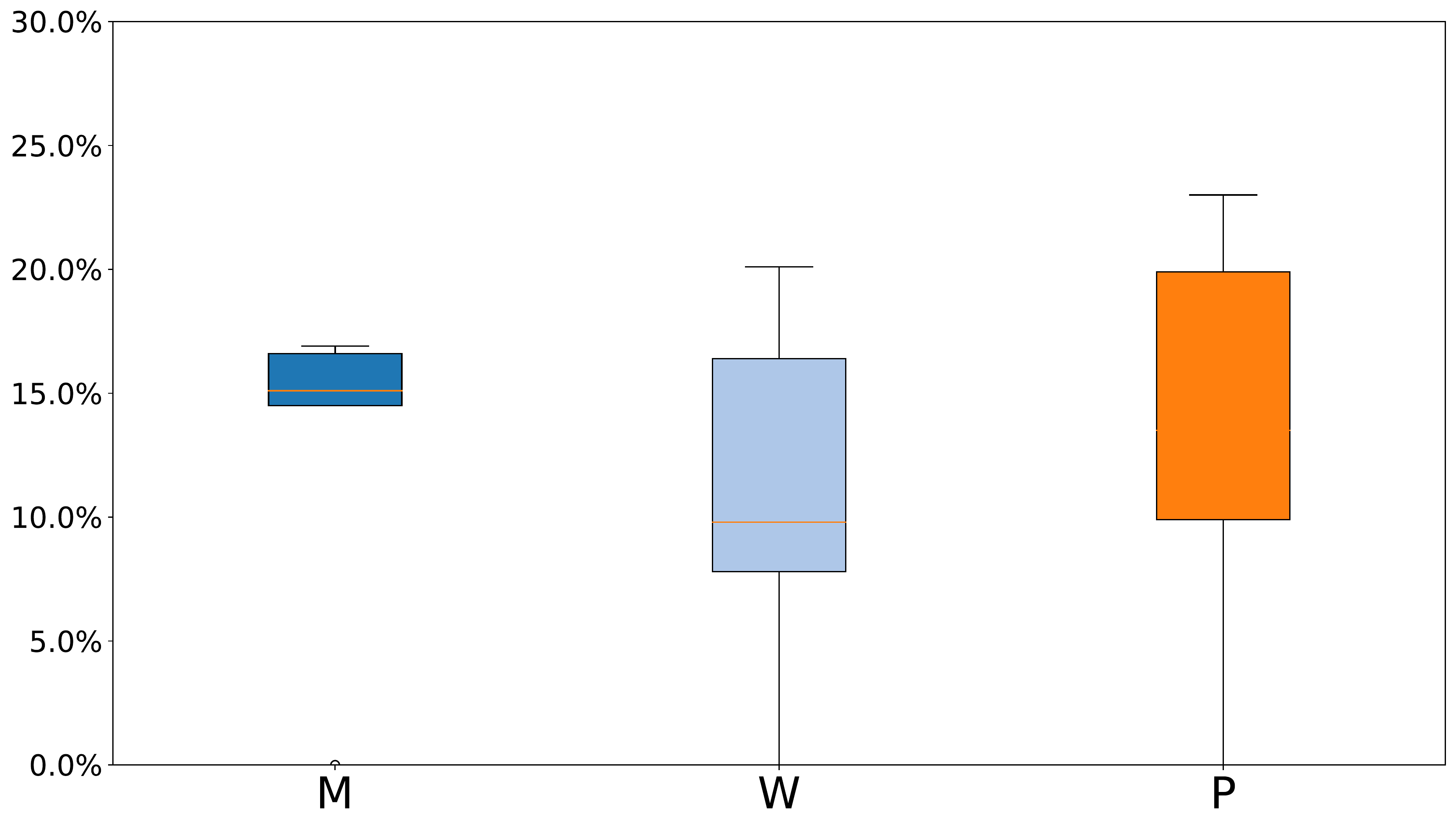}\end{minipage}}
	\hfill
	\subfigure[Branch Coverage (2h)]{\begin{minipage}[t]{0.23\linewidth}\centering\includegraphics[width=\linewidth]{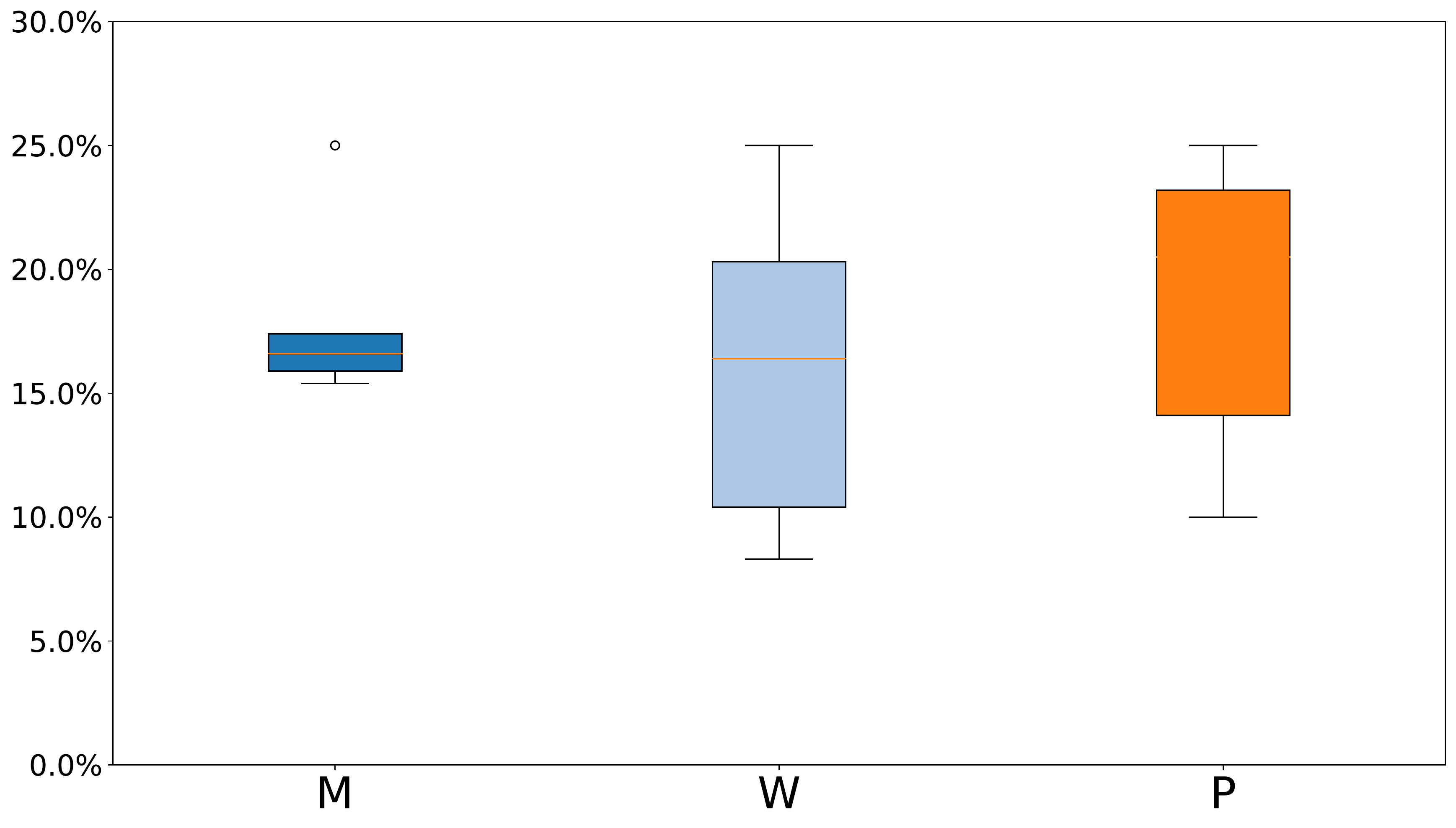}\end{minipage}}
	\hfill
	\subfigure[Branch Coverage (3h)]{\begin{minipage}[t]{0.23\linewidth}\centering\includegraphics[width=\linewidth]{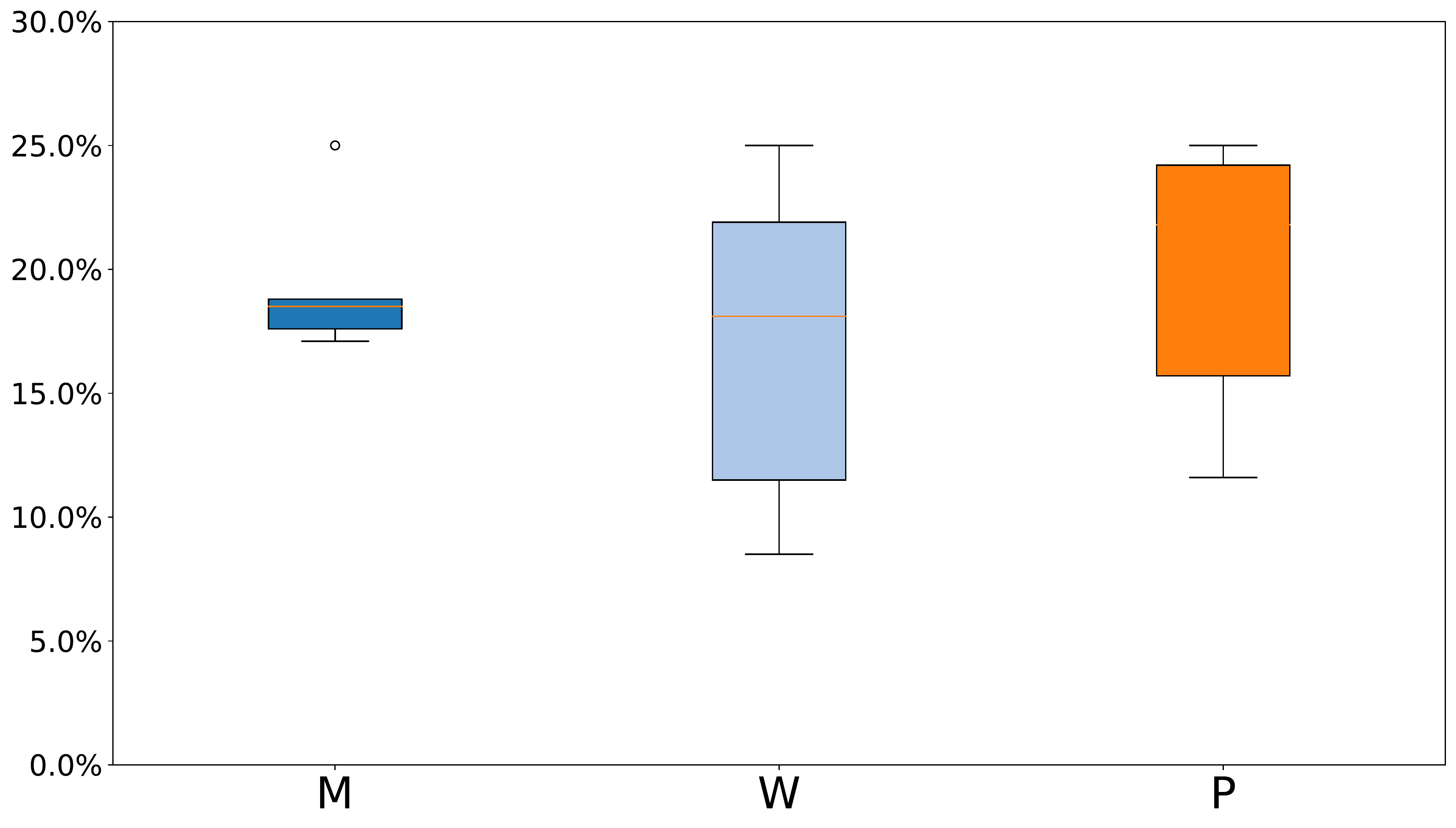}\end{minipage}}
	
	\subfigure[Detected Bug (500step)]{\begin{minipage}[t]{0.23\linewidth}\centering\includegraphics[width=\linewidth]{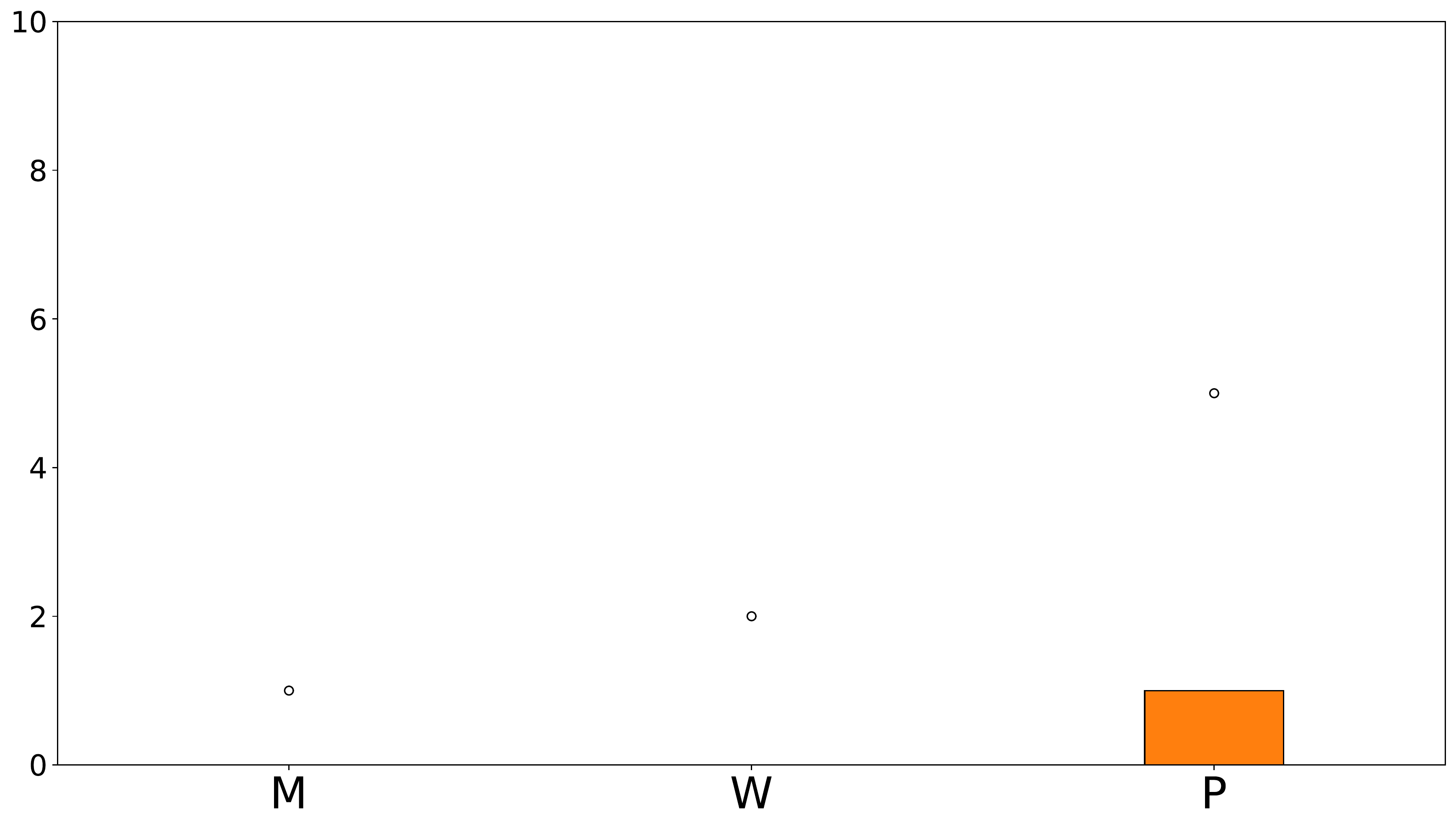}\end{minipage}}
	\hfill
	\subfigure[Detected Bug (1.5h)]{\begin{minipage}[t]{0.23\linewidth}\centering\includegraphics[width=\linewidth]{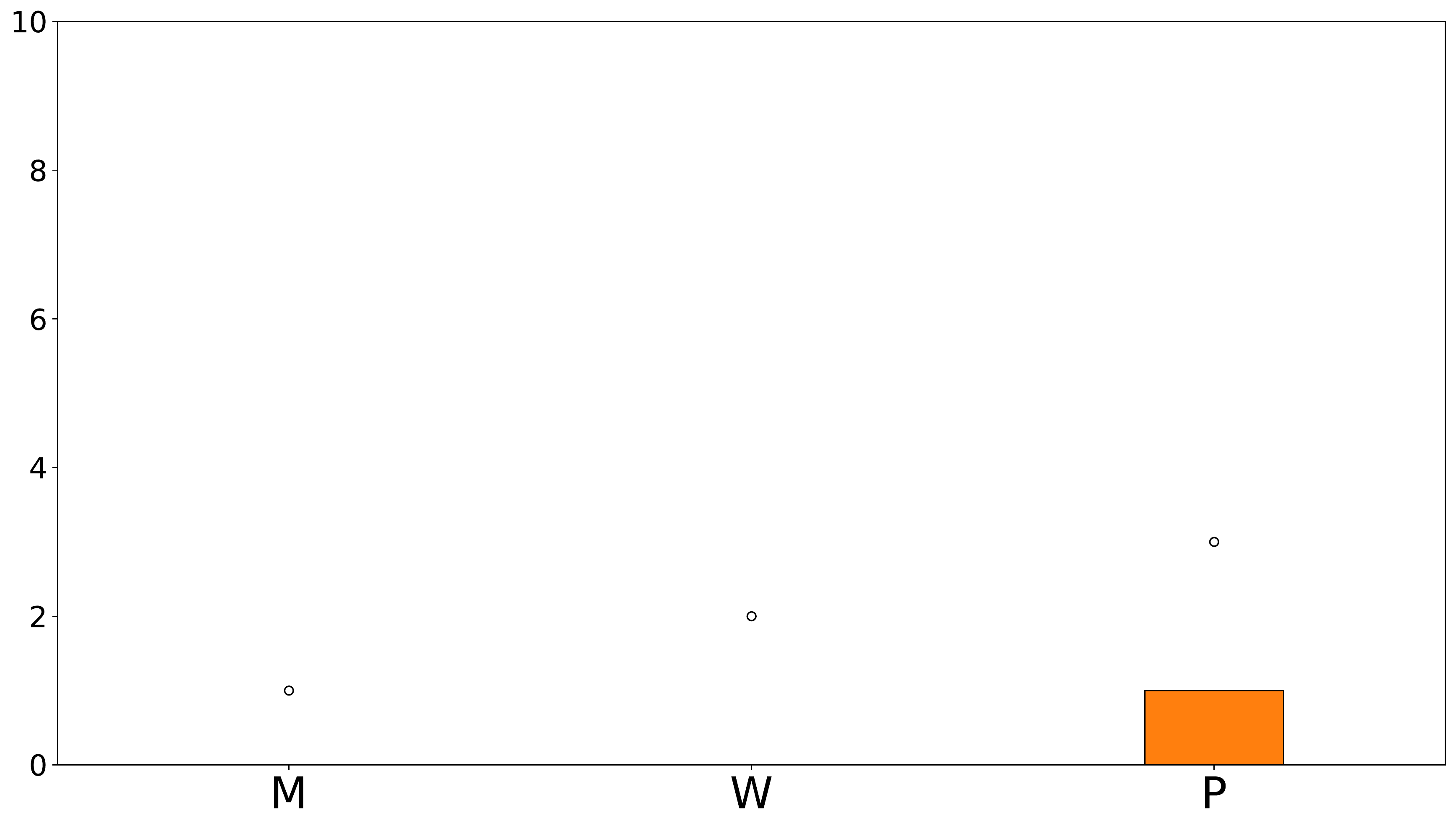}\end{minipage}}
	\hfill
	\subfigure[Detected Bug (2h)]{\begin{minipage}[t]{0.23\linewidth}\centering\includegraphics[width=\linewidth]{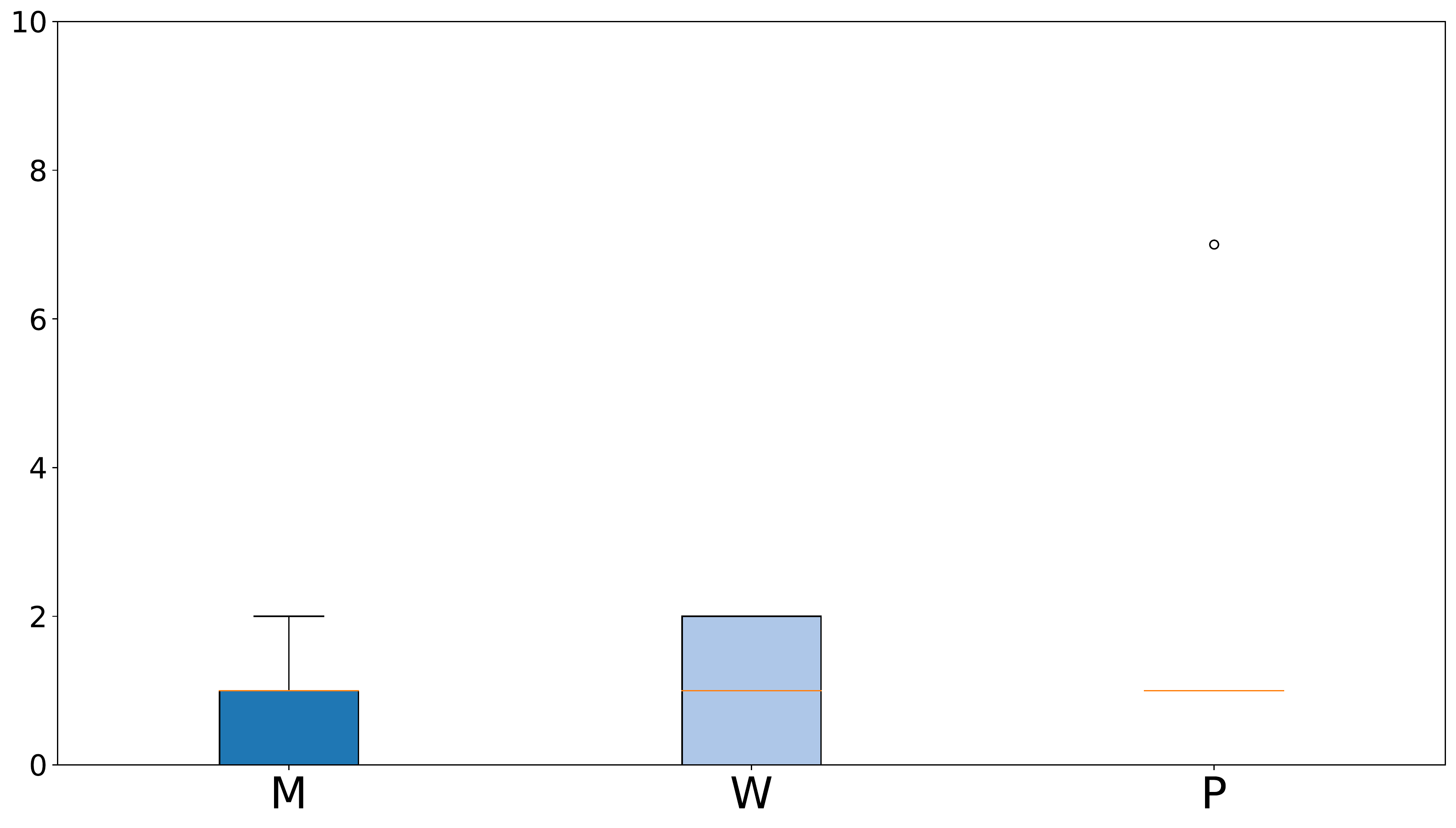}\end{minipage}}
	\hfill
	\subfigure[Detected Bug (3h)]{\begin{minipage}[t]{0.23\linewidth}\centering\includegraphics[width=\linewidth]{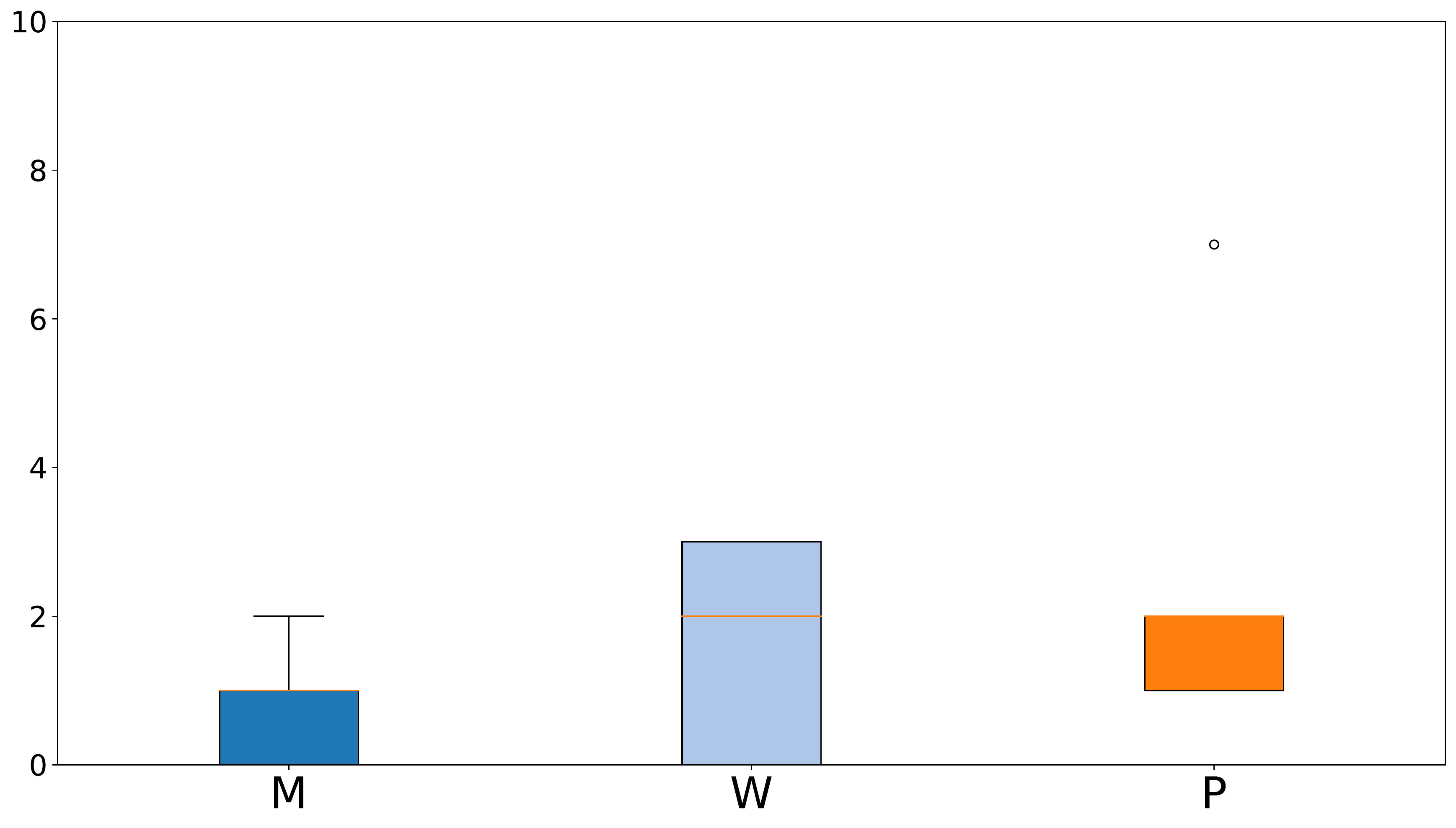}\end{minipage}}
	\centering
	\caption{Code Coverage \& Bug Detection Capability Comparison on Web Apps (M: \monkey, W: \webexplor, P: \toolname)}
\label{fig:expweb}
\end{figure}

The results on code coverage are presented in \tabref{tab:expsum}, \figref{fig:expmob} and \figref{fig:expweb}. According to the results, \toolname generally performs better than the baselines with regard to code coverage on both mobile apps and web apps. On mobile apps, \toolname achieves 7.2\% - 40.7\% more line coverage compared with different baselines after running 1.5 hours, 11.3\% - 42.8\% more line coverage after running 2 hours, and 11.4\% - 42.2\% more line coverage after running 3 hours. It achieves 4.4\% - 41.4\% more branch coverage compared with different baselines after running 1.5 hours, 4.4\% - 22.5\% more branch coverage after running 2 hours, and 7.9\% - 26.3\% more branch coverage after running 3 hours. On web apps, \toolname achieves 3.8\% and 15.5\% more line coverage compared with \monkey and \webexplor after running 1.5 hours, 0.4\% and 12.5\% more line coverage after running 2 hours, and 13.8\% and 16.6\% more line coverage after running 3 hours. It achieves 5.0\% and 22.7\% more branch coverage than the baselines after running 1.5 hours, 2.7\% and 15.5\% more branch coverage after running 2 hours, and 1.3\% - 15.7\% after running 3 hours.

The results indicate that \toolname can generate much fewer non-effective test events, while most test events generated by \monkey are non-effective, \ie on blank areas. Learning-based approaches are more effective than random and model-based approaches due to the exploration strategy design. Besides, the baseline approach miss many valid GUI widgets that do not appear in the layout XML files, which is also how \toolname generates more effective test events based on the widget identification from the visual perspective. Regarding the test generation efficiency, learning-based approaches, including \toolname, have extra overhead in processing GUI pages and exploring with RL frameworks. This is also the reason why \monkey generates more test events, though most of which non-effective. Generally, learning-based approaches, including LLM-based ones, are better than non-learning-based approaches.

\begin{figure}[!htbp]
  \subfigure[Mobile App]{
    \begin{minipage}[t]{0.48\linewidth}
      \label{fig:actmob}
      \centering
      \includegraphics[width=\linewidth]{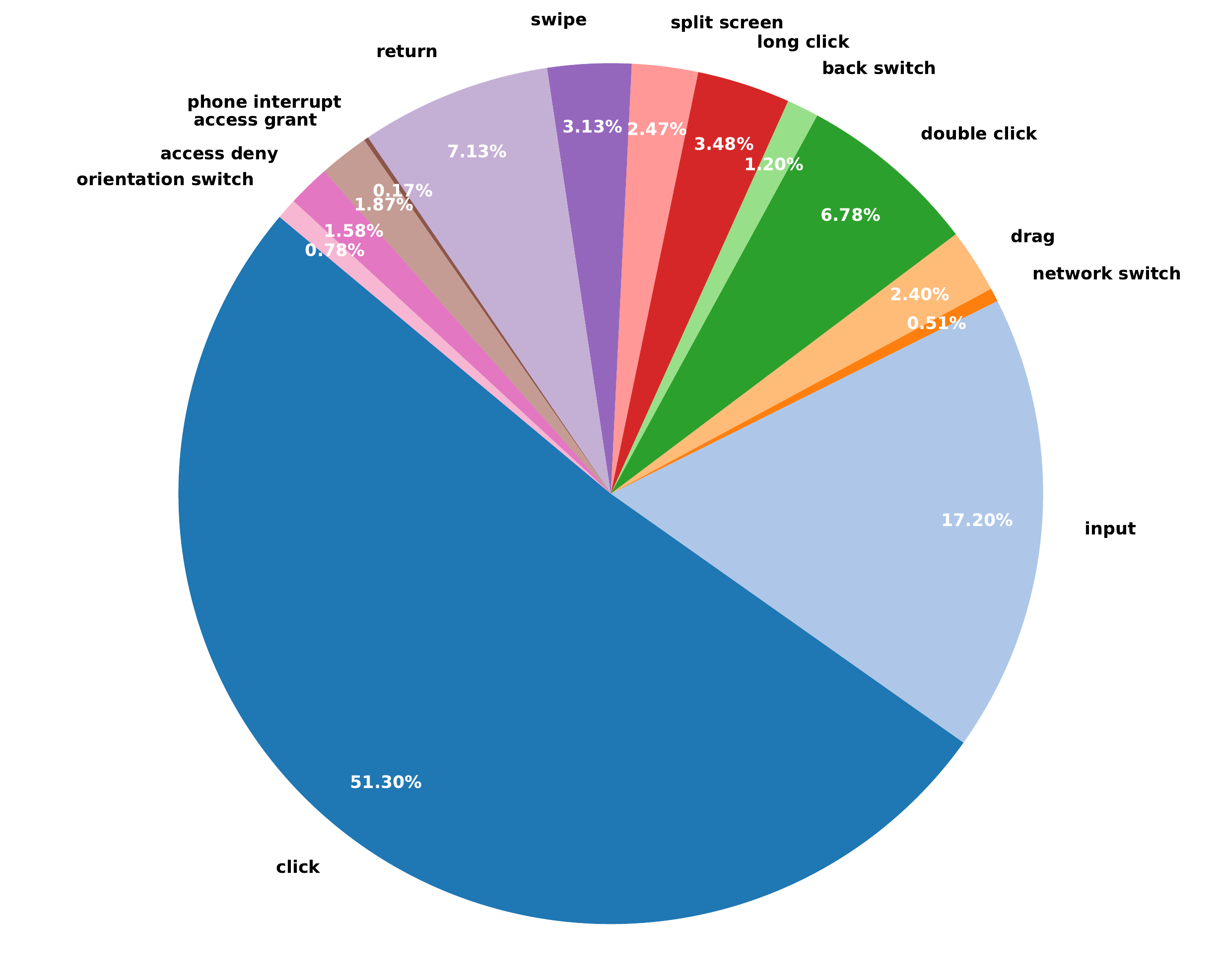}
    \end{minipage}
  }
  \hfill
  \subfigure[Web App]{
    \begin{minipage}[t]{0.48\linewidth}
      \label{fig:actweb}
      \centering
      \includegraphics[width=\linewidth]{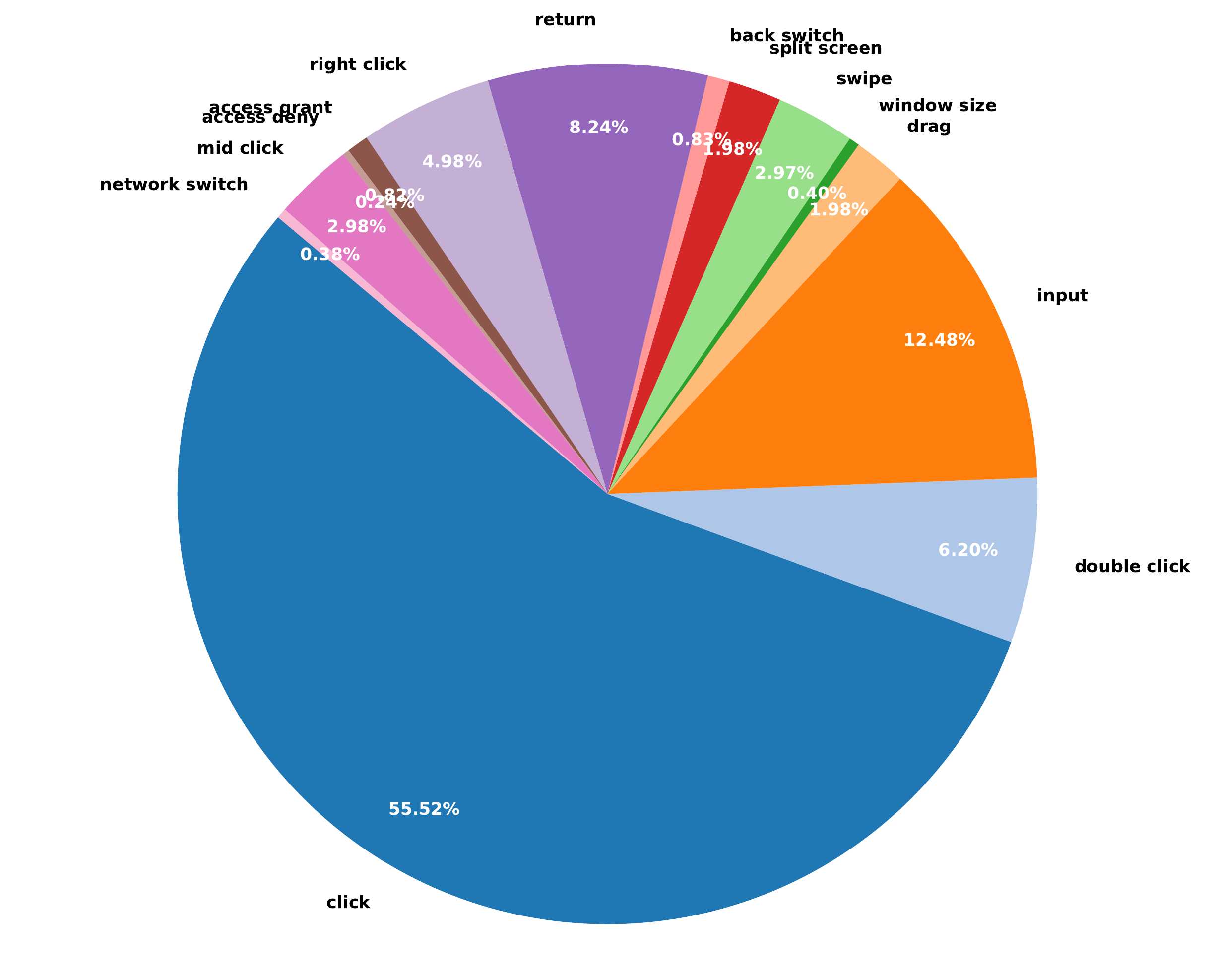}
    \end{minipage}
  }
  \centering
  \caption{Action Distribution in All Experiment Subjects}
\label{fig:expact}
\end{figure}

From the time cost aspect, we can find that the advantage of \toolname improves from 1.5-hour running to 3-hour running, we go through the test events and find that \toolname can generate longer test event sequences. For example, in the ``OmniNotes'' app, most functions are available only after logging in, while the entrance to login is not directly presented on the launching page, and a relatively long test event sequence is necessary to reach the login page, and then the exploration will continue after the login events. Given more time (1.5 hours to 2 hours, and to 3 hours), \toolname are more likely to generate longer test event sequences. This is also the reason that \toolname performs not so well as baselines on several apps, where the advantage of \toolname cannot be fully utilized.

Due to the test generation efficiency differences, The approaches are compared at the same test event quantity granularity. \toolname achieves 10.5\% - 41.0\% more line coverage compared with baselines on mobile apps. It achieves 4.1\% - 41.0\% more branch coverage than the baselines. On web apps, \toolname achieves 44.1\% and 1.5\% more line coverage compared with \monkey and \webexplor. It achieves 51.1\% and 2.8\% more branch coverage than the baselines. It is observed that the advantage of \toolname is more highlighted.

We investigate the actual generated test actions in our experiment, and the results are shown in \figref{fig:expact}. We can find that for both mobile and web apps, the ``click'' action takes over half of all generated actions, the reason is that most widgets are applicable for the ``click'' action. Except the ``click'' action, ``input'', ``return'', and ``double click'' are top three action types for both mobile and web apps. Other kinds of actions take a small percentage in all generated test events, which corresponds to the real app usage.

\begin{table}[!htbp]
\centering
\caption{Cross Coverage \& Bug Detection Analysis (X refers to the baseline approaches, M: Monkey, S: Stoat, T: TimeMachine, C: ComboDroid, H: Humanoid, A: APE, Q: Q-testing, G: GPTDroid, W: \webexplor, P: \toolname.}
\scalebox{0.8}{
\begin{tabular}{cc|cccccccc|cc}

\toprule
\multicolumn{2}{c|}{\multirow{2}{*}{Configuration}} & 
\multicolumn{8}{c|}{Mobile App} & 
\multicolumn{2}{c}{Web App} \\ \cmidrule{3-12}

\multicolumn{2}{c|}{} & M & S & T & C & H & A & Q & G & M & W \\ \midrule

\multirow{2}{*}{Line Coverage (500step)}   
& P $\uparrow$ X & 19.6\% & 27.6\% & 12.8\% & 22.1\% & 19.5\% & 12.8\% & 20.1\% & 19.3\% & 4.3\% & 12.0\% \\
& X $\uparrow$ P & 5.3\%  & 9.8\%  & 9.4\%  & 8.5\%  & 7.8\%  & 10.7\% & 7.3\%  & 7.1\%  & 2.7\% & 1.8\%  \\ \cmidrule{1-12}

\multirow{2}{*}{Line Coverage (1.5h)}      
& P $\uparrow$ X & 24.2\% & 24.6\% & 20.0\% & 23.4\% & 28.6\% & 28.3\% & 13.7\% & 13.6\% & 16.0\% & 36.7\% \\
& X $\uparrow$ P & 5.5\%  & 6.6\%  & 8.5\%  & 10.0\% & 8.1\%  & 11.2\% & 6.7\%  & 6.2\%  & 20.0\% & 0.7\%  \\ \cmidrule{1-12}

\multirow{2}{*}{Line Coverage (2h)}        
& P $\uparrow$ X & 20.0\% & 30.2\% & 15.5\% & 17.7\% & 15.4\% & 8.3\%  & 22.8\% & 15.8\% & 12.7\% & 16.6\% \\
& X $\uparrow$ P & 2.6\%  & 9.3\%  & 5.6\%  & 9.2\%  & 6.3\%  & 14.5\% & 11.6\% & 11.2\% & 5.8\% & 3.4\%  \\ \cmidrule{1-12}

\multirow{2}{*}{Line Coverage (3h)}        
& P $\uparrow$ X & 27.7\% & 29.6\% & 16.3\% & 25.4\% & 27.4\% & 31.5\% & 15.3\% & 12.9\% & 16.2\% & 39.2\% \\
& X $\uparrow$ P & 2.8\%  & 8.9\%  & 4.9\%  & 8.3\%  & 6.1\%  & 11.6\% & 3.0\%  & 3.0\%  & 6.8\% & 4.7\% \\ \cmidrule{1-12}

\multirow{2}{*}{Branch Coverage (500step)} 
& P $\uparrow$ X & 27.1\% & 53.7\% & 51.5\% & 52.4\% & 54.3\% & 56.8\% & 35.4\% & 34.2\% & 22.3\% & 24.2\% \\
& X $\uparrow$ P & 9.6\%  & 35.7\% & 14.8\% & 21.1\% & 14.5\% & 12.7\% & 21.5\% & 19.8\% & 21.3\% & 7.3\% \\ \cmidrule{1-12}

\multirow{2}{*}{Branch Coverage (1.5h)}    
& P $\uparrow$ X & 43.1\% & 21.8\% & 33.5\% & 32.5\% & 39.0\% & 37.6\% & 33.7\% & 29.2\% & 18.8\% & 49.6\% \\
& X $\uparrow$ P & 12.6\% & 21.9\% & 11.9\% & 16.1\% & 12.1\% & 11.0\% & 14.5\% & 12.9\% & 26.4\% & 14.9\% \\ \cmidrule{1-12}

\multirow{2}{*}{Branch Coverage (2h)}      
& P $\uparrow$ X & 22.3\% & 52.1\% & 54.5\% & 55.1\% & 55.6\% & 57.8\% & 57.3\% & 55.9\% & 26.7\% & 22.3\% \\
& X $\uparrow$ P & 7.7\%  & 8.8\%  & 12.7\% & 15.7\% & 13.3\% & 14.5\% & 29.2\% & 27.3\% & 19.3\% & 6.7\% \\ \cmidrule{1-12}

\multirow{2}{*}{Branch Coverage (3h)}      
& P $\uparrow$ X & 43.1\% & 23.2\% & 25.6\% & 27.2\% & 31.7\% & 35.2\% & 31.3\% & 27.5\% & 22.7\% & 48.9\% \\
& X $\uparrow$ P & 15.7\% & 15.7\% & 8.2\%  & 14.5\% & 10.7\% & 11.5\% & 18.5\% & 17.1\% & 14.1\% & 13.8\% \\ \cmidrule{1-12}

\multirow{3}{*}{Detected Bug (500step)}           
& P $\uparrow$ X & 102 & 102 & 100 & 96  & 94  & 89  & 97 & 96 & 5 & 4 \\
& P $\cap$ X     & 15  & 15  & 17  & 21  & 23  & 28  & 20 & 21 & 1 & 2 \\
& X $\uparrow$ P & 5   & 37  & 35  & 29  & 20  & 16  & 29 & 76 & 0 & 0 \\ \cmidrule{1-12}

\multirow{3}{*}{Detected Bug (1.5h)}
& P $\uparrow$ X & 97  & 106 & 103 & 100 & 93  & 92  & 55 & 90 & 3 & 2 \\
& P $\cap$ X     & 24  & 15  & 18  & 21  & 28  & 29  & 66 & 31 & 1 & 2 \\
& X $\uparrow$ P & 10  & 41  & 40  & 36  & 27  & 20  & 50 & 65 & 0 & 0 \\ \cmidrule{1-12}

\multirow{3}{*}{Detected Bug (2h)}
& P $\uparrow$ X & 109 & 110 & 101 & 108 & 109 & 101 & 87 & 71 & 8 & 7 \\
& P $\cap$ X     & 23  & 22  & 31  & 24  & 23  & 31  & 45 & 61 & 3 & 4 \\
& X $\uparrow$ P & 22  & 43  & 37  & 43  & 42  & 25  & 82 & 44 & 1 & 1 \\ \cmidrule{1-12}

\multirow{3}{*}{Detected Bug (3h)}
& P $\uparrow$ X & 114 & 122 & 107 & 110 & 116 & 114 & 80 & 96 & 9 & 7 \\
& P $\cap$ X     & 31  & 23  & 38  & 35  & 29  & 31  & 65 & 49 & 4 & 6 \\
& X $\uparrow$ P & 31  & 55  & 37  & 41  & 45  & 37  & 71 & 64 & 0 & 2 \\
\bottomrule  
        
\end{tabular}}
\label{tab:cross}
\end{table}

To further illustrate the advantage of \toolname, we conduct the cross analysis on code coverage. The results are shown in \tabref{tab:cross}. 

For the 500-step results on mobile apps, the baselines can only cover 5.3\% - 10.7\% more code lines and 9.6\% - 35.7\% more code branches than \toolname. However, \toolname can cover 12.8\% - 27.6\% more code lines and 27.1\% - 56.8\% more code branches than baselines in contrast. For the 1.5-hour results on mobile apps, the baselines can only cover 5.5\% - 11.2\% more code lines and 11.0\% - 21.9\% more code branches than \toolname. However, \toolname can cover 13.6\% - 28.6\% more code lines and 21.8\% - 43.1\% more code branches than baselines in contrast. For the 500-step results on web apps, the baselines can only cover 1.8\% - 2.7\% more code lines and 7.3\% - 21.3\% more code branches than \toolname. However, \toolname can cover 4.3\% - 12.0\% more code lines and 22.3\% - 24.2\% more code branches that are not covered by the baselines. For the 1.5-hour results on web apps, the baselines can only cover 0.7\% - 20.0\% more code lines and 14.9\% - 26.4\% more code branches than \toolname. However, \toolname can cover 22.3\% - 24.2\% more code lines and 18.8\% - 49.6\% more code branches that are not covered by the baselines. The data is similar for the 2-hour and 3-hour results, and \toolname can actually cover much different code that cannot be covered by the baselines. Although the advantage of \toolname in absolute coverage information might not be quite outstanding, the results on cross-coverage analysis show that \toolname can cover much code that cannot be covered by the state-of-the-art baseline approaches, which are obvious on both mobile apps and web apps. The results elaborate on the excellent performance of \toolname in covering more app source code.

\subsection{RQ2: Bug Detection}

In addition to code coverage, bug detection capability is more direct evidence to show the effectiveness of \toolname. The results in \tabref{tab:expsum}, \figref{fig:expmob} and \figref{fig:expweb} indicate that \toolname can detect more bugs compared with the baseline approaches. 

For the bug detection capability, \toolname detects 5 - 87, 5 - 87, and 9 - 83 more bugs than the baselines within 1.5 hours, 2 hours, and 3 hours, respectively, on mobile apps. Besides, \toolname detect 20 - 97 more bugs than the baselines within 500 test events on mobile apps. \toolname detects 2 - 3, 6 - 7, and 5 - 9 more bugs than the baselines within 1.5 hours, 2 hours, and 3 hours, respectively, on web apps. \toolname detects 4 - 5 more bugs within 500 test events on web apps. We review the detected bugs and confirm that all the bugs are crash bugs. We have reported the bugs detected by \toolname to the developers, and 21 bugs are confirmed or fixed.

The cross analysis is conducted on the detected bugs (\tabref{tab:cross}). For mobile apps, \toolname detects 55 - 106, 71 - 110, and 80 - 122 bugs that the baselines cannot detect within 1.5 hours running, 2 hours running, and 3 hours running, respectively. \toolname detects 89 - 102 bugs over baselines within 500 test events. For web apps, \toolname detects 2 - 3, 3 - 4, and 7 - 9 bugs that the baselines cannot detect within 1.5 hours running, 2 hours running, and 3 hours running, respectively. \toolname detects 4 - 5 bugs over baselines within 500 test events. The bugs uniquely detected by \toolname are much more than the bugs uniquely detected by baselines. The results indicate that \toolname can detect more bugs that cannot be detected by the baseline approaches. The results come from the design of GUI state abstraction, which help better characterize the GUI and corresponding widgets to encourage effective exploration. Another noteworthy observation is that \toolname can detect more bugs with higher efficiency (within 500 test events), especially compared with \qtesting on mobile apps.

Generally speaking, the correlation relationship between the code coverage and bug detection capability is positive based on the experiment results of different time costs (1.5 hours, 2 hours, and 3 hours), which means that the more code one approach can cover, the more likely it can detect more bugs. However, the correlation relationship is also affects by other factors. For example, the distribution of bugs within the whole app is unknown, and it is hard to assign specific targets to automated GUI testing approaches to detect bugs. Second, different inputs can lead to different. Due to the weak correlation relationship between the code coverage and bug detection capability, we present the two metrics to better illustrate the performance of \toolname.

The data show that \toolname has an excellent performance in detecting bugs on both mobile and web platforms. Specifically, the cross analysis on detected bugs proves that \toolname is capable of detecting bugs that are hard to be detected by the baseline approaches. We have an in-depth analysis in the following section.

\subsection{RQ3: Advantage Analysis}

We have an in-depth investigation of the results of \toolname over the baselines, and we list three of the significant reasons that help improve the effectiveness of \toolname. 

\textbf{Widget Granularity Exploration.} The widget-based RL exploration is one of the most significant novel contributions of this paper. Instead of retrieving the action targets from the layout XML files like some existing approaches, \toolname directly starts from the GUI page and obtains more effective widgets to alleviate the problems presented in \secref{sec:challenge}. This is one of the significant reasons for the success of \toolname. However, the widget accuracy \cite{chen2020object} is one of the obstacles that hinder \toolname from achieving a better performance.

\textbf{Q-network with State and Action Embedding.} After getting the effective embedding to the states and the actions, the Q-value of one specific state-action pair will be refreshed, so the GUI pages with similar appearances that carry similar functions will be recognized to avoid repeatedly covering the same code snippets. The design of the embedding and the Q-network gives \toolname the capability of processing the huge exploration state and action space. However, there exist a few situations where the assumption that similar GUI layouts are similar pages may not work, so the Q-network may under-estimate the values of such pages.

\textbf{Long Semantic Action Sequence.} This point is the most significant one in our opinion, which is that \toolname shows the potential of generating a long action sequence within the context. During the generation of long semantic action sequences, it may be interrupted if some irrelevant actions are generated. Given that the ``future'' reward is considered in the value estimation of current actions, a longer action sequence should be preferred by the RL exploration mechanism, and such a long action sequence tends to have context semantics from the perspective of human testers.

In this paper, we implement the \toolname and conduct an empirical evaluation on Android and web platforms. However, we hold that \toolname is a platform-independent approach, and will be effective in all GUI-based platforms, including iOS, harmonyOS, desktop operating systems, or even some embedded systems. The only accordance of \toolname is the analysis of the widget extraction and the layout characterization. As long as the platform has a GUI-based user interface, \toolname will be capable of exploring the state and action space. We believe \toolname has strong generalizability on different platforms and behaves at zero cost when adapting different platforms.

Another point is about efficiency. \toolname seems to have relatively poor efficiency compared with random-based approaches, \ie \monkey. However, this is the common problem of model-based approaches (based on dynamic analysis) or learning-based approaches. As far as we know, none of the existing work has a comparison of the action generation efficiency. We hold that in the future, \toolname will be a complement to the random-based approaches, and will cover more code that cannot be covered by the existing approaches.

\subsection{Threats to Validity}

The main threat may be the \textbf{app selection}. In order to reduce this threat, we use the app under experiment from the existing research, including \cite{mao2016sapienz, su2017guided, pan2020reinforcement, zheng2021automatic}. We use widely-used open-source apps from a popular list \cite{liu2020androzooopen} because some apps from the above papers are not accessible due to maintenance problems. Besides, the used apps cover many different categories, which expand the scenarios of GUI layout styles, ensuring the generalizability of \toolname.

Another threat may be the \textbf{parameter or hyperparameter settings}, like the thresholds in the widget recognition algorithm, or the deep learning model training. In order to alleviate the bias, we set the parameters according to domain knowledge or following the common practice from existing work. The training data are fully isolated from the experiment subjects we use in the evaluation. For some critical parameters, we conduct pilot studies to find suitable settings before the evaluation. For the baselines, we use the default settings or the open-source package to avoid possible biases.

\toolname contains several deep learning models, and \textbf{the performance of the proposed deep learning networks} may be a potential threat. In order to mitigate this threat, we try our best to collect more data to train the neural networks. We use networks with simple structures to avoid performance problems that may affect the effectiveness of \toolname.

\subsection{Discussion}

Compared with the baselines, \toolname has achieved promising results. However, there are some limitations for the proposed approach, which can also be solved in future work. First, due to the special design of app GUI state abstraction, \toolname can only process apps with table-like or list-like layout. Although such layout takes a vast majority percentage of all kinds of apps, there are few categories of apps \toolname cannot process, like the video games. However, with the development of vision-based GUI processing technologies, we believe the widget extraction and layout characterization of app GUI can be advanced. Second, although \toolname outperforms the baseline approaches, including the LLM-based tool, we believe it can be more effective to combine both, taking advantages of the app understanding capability of LLMs and exploration reward mechanism of RL frameworks. Third, although \toolname shows preliminary capabilities of generating long semantic action sequences, we believe the potentials of combining app business logic is not fully explored. Integrating the domain knowledge of human testers may further advance the capability of RL frameworks.

\section{Related Work}

\subsection{Automated Software Exploration}

The most basic strategy of automated software testing is the random-based strategy. Among them, \monkey is the most widely used tool \cite{google2022monkey}, and performs quite well on specific benchmark apps \cite{choudhary2015automated}. \monkey is capable of generating a large number of test events with high efficiency. However, the shortcomings are obvious. Due to the lack of effective guidance, \monkey generates a large percentage of noneffective or redundant test events, posing a threat to the testing effectiveness. Such a problem is alleviated by restricting the GUI states \cite{daragh2021deep}, but the exploration process still lacks guidance.

Approaches adopting model-based strategy \cite{amalfitano2014mobiguitar, yu2015incremental} build specific models for the testing exploration with static or dynamic (or combination) app analysis \cite{mesbah2011invariant}. 
AndroidRipper \cite{amalfitano2012using} proposed by Amalfitano \etal adopts a user-interface driven ripper to explore the app GUI to exercise the app in a structured way. 
Baek \etal \cite{baek2016automated} present the GUICC, which provides the selection of multiple abstraction levels for GUI model generation with a set of multi-level GUI comparison criteria. 
Aimdroid \cite{gu2017aimdroid} introduced by Gu \etal is a tool that aims to manage the exploration of activities and to minimize unnecessary transitions between them through an activity-insulated multi-level strategy during the testing. 
Biagiola \etal propose SubWeb \cite{biagiola2017search}, which takes advantage of the navigation structure implicitly specified by developers for web testing. 
DIG \cite{biagiola2019diversity} pre-selects the most promising candidate test cases based on their diversity from previous tests. GoalExplorer \cite{lai2019goal} firstly statically models the app GUI and transitions, and then guides the dynamic exploration of the app to the particular target of interest. 
Sapienz \cite{mao2016sapienz}, proposed by Mao \etal, is a typical tool that uses multi-objective search-based testing to automatically explore and optimize test sequences. 
Stout \cite{su2017guided} is a representative and representative model-based tool, which uses a stochastic Finite State Machine model to describe the behavior of AUT. Due to the limitation of the modeling algorithms, the model-based approach can neither precisely nor completely characterize the apps under test.

Benefiting from the development of deep learning and machine learning models, learning-based automated testing technologies also emerge. 
White \etal \cite{white2019improving} propose a machine learning-based technique to improve GUI testing by automatically identifying GUI widgets in screenshots.
Li \etal \cite{li2019humanoid} propose Humanoid, an automated black-box Android app testing tool based on deep learning models.
Reinforcement learning is one group of suitable algorithms for app exploration.
Koroglu \etal \cite{koroglu2018qbe} propose a fully automated black-box testing methodology, which explores GUI actions using Q-Learning.
Zhao \etal \cite{zhao2022dinodroid} propose DinoDroid, an approach based on deep Q-networks to automate testing of Android apps. 
Lv \etal \cite{lv2022fastbot2} introduce an automated model-based GUI testing technique to accelerate the testing process and experiments on two popular industrial apps show its effectiveness.
Wuji \cite{zheng2019wuji} is a typical approach that utilizes RL algorithm to explore video game apps. It balances winning the game and exploring the space of the game to uncover more bugs. 
\qtesting is an effective tool for Android testing \cite{pan2020reinforcement} using the RL algorithm, it trains a neural network to divide different states at the granularity of functional scenarios. 
\webexplor \cite{zheng2021automatic} is an advanced tool for web app testing with the RL algorithm. It adopts a curiosity-driven model to generate high-quality test events containing temporal logical relations. 
Romdhana \etal \cite{romdhana2021deep} compare different RL algorithms in mobile app testing. 

Among the aforementioned automated testing approaches, much progress is made to improve the automated software testing effectiveness and efficiency. However, as claimed in \secref{sec:challenge}, current approaches are still faced with critical problems, which hinder the approaches from showing better performance. Besides, as far as we know, none of the current approaches conduct platform-independent app testing, which builds a high bar for app developers when faced with the sharply growing number of running platforms.

\subsection{App GUI Understanding}

Due to the fragmentation problem of apps \cite{wei2016taming}, more and more work starts to apply GUI understanding technologies to assist the software engineering tasks \cite{yu2021layout}. 
One group of researchers uses GUI understanding technologies to conduct reverse engineering and generate code snippets from GUI images \cite{chen2018ui, moran2018machine, nguyen2015reverse, zhao2021guigan}. 
Such tools adopt traditional computer vision technologies or deep learning/machine learning models to analyze the GUI images.

Another important direction of using GUI understanding in software testing is test report analysis and optimization. 
Yu \etal \cite{yu2019crowdsourced} use image understanding technologies to generate reports for app screenshots indicating bugs. 
Yu \etal \cite{yu2021prioritize} prioritize crowdsourced test reports with deep image and text semantic understanding. 
Besides, representative work like \cite{feng2015test, feng2016multi, wang2019images} all use GUI image understanding in crowdsourced test report optimization. 
Studies \cite{bernal2020translating, feng2022gifdroid} have been proposed to help reproduce the video-based test reports by analyzing the GUI information in the video test reports. 
Tango \cite{cooper2021takes} is capable of detecting duplication in video-based test reports.
Also, GUI understanding is widely used in record and replay test scripts. 
Behrang \etal \cite{behrang2018test} introduce AppTestMigrator to migrate test cases between apps using the similarity among GUI widgets. 
Qian \etal \cite{qian2020roscript} present a tool that uses CV algorithms to recognize GUI widgets and then controls robots to complete the automated testing tasks. 
Yu \etal \cite{yu2021layout} realize the cross-platform record and replay of test scripts with the image and layout characterization.
One significant basis for applying GUI understanding technologies in software testing is widget identification. 
Chen \etal discuss the widget recognition algorithms in \cite{chen2020object}. 
Xiao \etal propose IconIntent \cite{xiao2019iconintent} to understand the intents of GUI widgets to identify sensitivity. 
Liu \etal propose a tool to help detect UI display issues with deep GUI understanding \cite{liu2020owl}. 
Chen \etal \cite{chen2020unblind} propose a deep learning model to help predict the labels of specific GUI widgets. 
Chen \etal \cite{chen2020wireframe} introduce a GUI design search engine to assist GUI designers.

The above studies inspire us to start automated testing from the GUI perspective with computer vision technologies. Moreover, considering the nature that some special widgets, like embedded \texttt{Canvas} elements or other customized elements, or the embedded HTML pages, cannot be retrieved from GUI layout files with traditional widget identification approaches, the visual analysis on app GUI improves the effectiveness of automated software testing.

\section{Conclusion}

To tackle the challenges of automated explorative testing, this paper proposes an effective, platform-independent GUI testing framework with image embedding and reinforcement learning. \toolname integrates a novel algorithm to embed the app GUI for the state comparison in the RL algorithm and uses a deep neural network to construct the Q-network for the state-action value determination. \toolname is the first approach that bridges the gap of app exploration on different platforms. The experiment results show that \toolname can cover 6.3--41.4\% more code on mobile apps and 1.5--51.1\% more code on web apps, with much of the code not covered by the baselines, and it can detect 128 unique bugs on mobile and web apps, including ~100 ones that are not detected by the baselines. \toolname is the first approach to independently test apps of different platforms with excellent effectiveness.

\begin{acks}
The authors would like to thank the editors and anonymous reviewers for their time and comments.
This work is supported partially by the National Natural Science Foundation of China (62141215, 61932012, 62372228),  and the Science, Technology and Innovation Commission of Shenzhen Municipality (CJGJZD20200617103001003).
\end{acks}

\bibliographystyle{ACM-Reference-Format}
\bibliography{main}

\end{document}